\documentclass[12pt]{iopart}
\usepackage{iopams} 
\usepackage{xcolor}
\usepackage{multicol}
\usepackage{multirow}
\usepackage{graphicx}
\usepackage{blindtext}
\usepackage{hyperref}
\usepackage{epstopdf}
\hypersetup{
    colorlinks=true,
    citecolor=black,
    linkcolor=black,
    filecolor=magenta,      
    urlcolor=black,
    }
\begin{document}

\title[]{Global Linear and Nonlinear Gyrokinetic Simulations of Tearing Modes}

\author{T.~Jitsuk$^1$, A.~Di~Siena$^2$, M.J.~Pueschel$^{3,\,4,\,5}$, P.W.~Terry$^1$,\\ F.~Widmer$^{2,6}$, E.~Poli$^2$, and J.S.~Sarff$^1$}
\vspace{10pt}
\address{$^1$Department of Physics, University of Wisconsin-Madison, WI 53706, USA}
\address{$^2$Max-Planck-Institut für Plasmaphysik, D-85748 Garching, Germany}
\address{$^3$Dutch Institute for Fundamental Energy Research, 5612 AJ Eindhoven, \\~\,The Netherlands}
\address{$^4$Eindhoven University of Technology, 5600 MB Eindhoven, The Netherlands}
\address{$^5$Department of Physics \& Astronomy, Ruhr-Universit{\"a}t Bochum, D-44780 Bochum, Germany
}
\address{$^6$National Institutes of Natural Sciences, Headquarters for Co-Creation Strategy, Tokyo, Japan}
\ead{\url{jitsuk@wisc.edu}}
\vspace{10pt}
\begin{indented}
\item[]\today
\end{indented}

\begin{abstract}
To better understand the interaction of global tearing modes and microturbulence in the Madison Symmetric Torus (MST) reversed-field pinch (RFP), the global gyrokinetic code \textsc{Gene} is modified to describe global tearing mode instability via a shifted Maxwellian distribution consistent with experimental equilibria. The implementation of the shifted Maxwellian is tested and benchmarked by comparisons with different codes and models. Good agreement is obtained in code-code and code-theory comparisons. Linear stability of tearing modes of a non-reversed MST discharge is studied. A collisionality scan is performed to the lowest order unstable modes ($n=5$, $n=6$) and shown to behave consistently with theoretical scaling. The nonlinear evolution is simulated, and saturation is found to arise from mode coupling and transfer of energy from the most unstable tearing mode to small-scale stable modes mediated by the $m=2$ tearing mode. The work described herein lays the foundation for nonlinear simulation and analysis of the interaction of tearing modes and gyroradius-scale instabilities in RFP plasmas.
\end{abstract}

%\submitto{\NF}
%\maketitle

\section{Introduction}
~~~~~Tearing modes are a well-known fluctuation in toroidally-confined plasmas and can significantly influence the stability and performance of fusion devices. They produce magnetic islands, are associated with cyclical oscillations like sawteeth, drive disruptive events, and interact with turbulence which potentially leads to significant plasma transport and energy losses \cite{hornsby2015seed, ishizawa2019multi}. When tearing modes excite a turbulent cascade, fluctuation energy reaches small scales where it can impact micro-scale instabilities and fluctuations \cite{duff2018observation, williams2017turbulence}.
Extensive research has been dedicated to understanding the linear and nonlinear physics of tearing modes, and in developing effective strategies to mitigate their detrimental effects.

One particular toroidal device where tearing modes are active and play a fundamental role is the reversed-field pinch (RFP) \cite{ho1991nonlinear,sarff1994fluctuation,chapman1996sawteeth,chapman1997fluctuation,chapman2002high,chapman2009improved,marrelli2021reversed}. Tearing modes in RFPs are global, system-scale modes driven by the current profile across its radial domain. These modes grow to finite amplitude and exhibit nonlinear couplings with each other and with small-scale fluctuations. Experimental measurements and nonlinear magnetohydrodynamic (MHD) computations have provided detailed analyses, revealing that the coupling between tearing modes of different poloidal and toroidal mode numbers $m$ and $n$ leads to the conversion of poloidal magnetic flux in the plasma core region to toroidal magnetic flux near the reversal surface. This results in a self-generated toroidal flux, commonly referred to as the dynamo effect \cite{ho1991nonlinear, hokin1991global}. Moreover, interactions among multiple tearing modes resonant at different radii enhance energy and momentum transport \cite{fiksel1994measurement, ebrahimi2007momentum}. When the activity of tearing modes is reduced through control of the plasma current profile in pulsed-poloidal-current-drive (PPCD) experiments, plasma confinement improves, characterized by increasing plasma beta and electron temperature, as well as reduced electron heat diffusivity \cite{chapman2009improved}. RFPs can furthermore exhibit a plasma state called quasi-single-helicity (QSH), where the lowest resonant mode at the inner radii singly dominates the tearing-mode spectrum, leaving other, higher-resonant modes at larger radii insignificant. This state is similar to having multiple tearing modes suppressed, which limits nonlinear tearing interactions, resulting in less magnetic stochasticity, increased plasma current, and therefore enhanced confinement \cite{escande2000qausi, martin2003overview, lorenzini2009self}.

Nonlinear interactions involving tearing modes occur not only between global-scale modes but also across an extended range of scales, as directly observed and inferred in a number of experiments \cite{ren2011experimental, thuecks2017evidence, titus2021dissipation}. As tearing modes cascade to small scales, they continue to interact with MHD fluctuations. They can also interact with drift-wave-type microinstabilities and turbulence. Such interactions, which involve not just different scales but different physics, are often described as multi-scale interactions. An example of a multi-scale interaction was demonstrated in a gyrokinetic simulation of density-gradient-driven trapped-electron-mode ($\nabla n$-TEM) turbulence in the RFP with a representation of tearing mode activity in terms of a constant-in-time magnetic-field perturbation modeling residual tearing fluctuations. The simulations showed that the magnetic fluctuation suppressed the zonal flows generated nonlinearly in TEM turbulence, resulting in an increase of the electrostatic heat flux \cite{williams2017turbulence}. 

Similar effects stemming from large-scale magnetic perturbations on zonal-flow-regulated microturbulence have also been observed in the tokamak. Resonant magnetic perturbations (RMPs), which are externally applied to plasmas for the suppression of edge-localized modes \cite{evans2008rmp}, were found in DIII-D experiments to increase transport levels and lower energy confinement \cite{mckee2013increase}. Local gyrokinetic simulations using a \textit{ad-hoc} radial magnetic perturbation similar to that of RFP modeling were carried out for a DIII-D L-mode scenario. The simulations indicate that incremental increases in turbulent fluctuations are related to the reduction of zonal-flow levels by this perturbation \cite{williams2020impact}. Similarly, the interaction between microtearing and electron-scale turbulence is partly governed by zonal flows and can be regulated via the addition of RMPs \cite{pueschel2020multi}. Thus, the physical mechanism at play is largely the same as that in the RFP. This modeling approach is most appropriate for external perturbations like the RMP. It is an approximation for the effect of internal magnetic perturbations from collective modes, both because collective modes have their internally consistent linear and nonlinear dynamics and because external perturbations cannot capture any possible back-reaction of the microturbulence on the collective mode. This motivates the present, more realistic and accurate approach.

Although extensive investigations have been conducted to examine self-interacting tearing modes within the RFP, interactions simultaneously involving micro- and macro-scale fluctuations have received comparatively limited attention and are therefore less understood. The limited progress on multi-scale interactions to date is due, at least in part, to the fact that the self-consistent coupling of two distinct scales is difficult to treat computationally. While studies of single-scale tearing modes and nonlinear couplings of different helicities can be performed entirely within MHD models, coupling with microinstabilities requires models that can capture gyroradius-scale drift-wave physics. One class of drift-wave-capable models is gyrokinetics \cite{brizard2007foundations}, which generally provides a comprehensive representation of micro-scale instability physics. Efforts have also been made by employing the gyrokinetic model to study global MHD modes. One of the prior studies developed the particle-in-cell gyrokinetic toroidal code (GTC), coupling a fluid electron description with gyrokinetic ions, reproduced resistive tearing modes \cite{liu2014verification} and collisionless tearing modes \cite{liu2016verification}, and returned correct eigenmode structures and growth-rate scaling in comparison with MHD computations. Self-consistent modeling including kinetic electrons was developed in the continuous $\delta f$ code GKW, specifically for studying the stability of linear tearing modes in three-dimensional toroidal geometry \cite{hornsby2015linear} and to study the seed-island generation and nonlinear interaction of tearing modes and electromagnetic turbulence \cite{hornsby2015seed, hornsby2015non}. 

As already stated, previous gyrokinetic studies of multi-scale interactions in RFP plasmas imposed a constant-in-time external $A_\parallel$ as a simple model for a residual tearing fluctuation \cite{williams2017turbulence, carmody2015microturbulence}. However, tearing-scale instabilities in experiments are internally consistent collective modes driven by a radial gradient in the background current parallel to the mean magnetic field. To better model this situation, a self-consistent description is needed. To this end, Ref.~\cite{williams2019interactions} implemented a current-gradient drive as part of the fluctuating distribution function. The implementation, restricted to slab geometry, allowed for a sinusoidal current profile with adjustable magnitude and phase, which are able to drive a (slab) tearing mode. Several aspects of this approach impose limitations: local slab simulations do not allow for a realistic equilibrium, and linear solutions of the gyrokinetic model equations result in slab-specific intermediate-scale modes. As these are not present in the experiment, they interfere with the study of experimentally relevant multi-scale interactions. 

 To more realistically model tearing modes, a shifted Maxwellian equilibrium distribution is implemented in the global version of the gyrokinetic code \textsc{Gene} \cite{jenko2000electron, gorler2016intercode, di2018non, di2019implementation, genecode}, consistent with a specified safety factor profile. Details related to the implementation of the shifted Maxwellian distribution are presented in this work, including the equilibrium distribution with shifted velocity, its effect on the perturbed distribution and the instability driving terms, and its modification of velocity moments that enter into the calculation of the self-consistent fields. The implementation is then benchmarked against several plasma models, including a reduced fluid model and gyrokinetic codes based on both continuum $\delta f$ and particle-in-cell full-$f$ approaches. Of particular note is benchmarking with the ORB5 code \cite{jolliet2007global,lanti2020orb5}, which was also modified to incorporate a shifted Maxwellian equilibrium distribution for current-gradient-driven modes \cite{mishchenko2019pullback,Mishchenko2022gyrokinetic}. In preparation for multi-scale simulations, linear tearing modes in a non-reversed RFP discharge are obtained. Linear initial-value calculations for the fastest-growing mode show that the lowest resonant mode is dominant. At large $k_y$ the fluctuations become electrostatic and are identified as $\nabla n$-TEMs \cite{ernst2009role}. The nonlinear evolution of the RFP tearing mode is obtained and studied in detail for the first time in gyrokinetics. These studies provide the basis for carrying out multi-scale interactions of tearing modes with gyroradius-scale turbulence in future work. 

The remainder of the paper is organized as follows. Details of the numerical approach as well as the gyrokinetic treatment of the current gradient and its implementation are discussed in Sec.~2. Numerical benchmarking against a variety of published results is elaborated in Sec.~3. Linear and nonlinear tearing mode evolution in the RFP system are studied in Sec.~4, and a summary and discussion are provided in Sec.~5.
\section{Current-Gradient Implementation}
The gyrokinetic framework \cite{brizard2007foundations} allows for the evaluation of microinstability physics as well as nonlinear interactions between fluctuations across a range of scales. Here, the global version of the gyrokinetic turbulence code \textsc{Gene} \cite{jenko2000electron, gorler2016intercode, genecode},  is employed, which has been used with non-Maxwellian background distributions \cite{di2018non, Di_Siena_2018}. \textsc{Gene} solves a Vlasov-Maxwell system of equations with a $\delta f$ approach, in which the total distribution $f$ is split into the static background distribution $F_0$ and the perturbed distribution $\delta f$, such that $f=F_{0}+\delta f$, with self-consistent field equations. The gyrokinetic Vlasov equation for each plasma species $j$ is represented by 
\begin{equation}
    \frac{\partial f_j}{\partial t} =\mathcal{L}[f_j]+\mathcal{N}[f_j]. 
\end{equation}
In this equation, $\mathcal{L}$ is the linear operator, containing among other terms the instability drive due to profile gradients for small-scale drift-wave instabilities, and $\mathcal{N}$ is the electromagnetic version of the $\textbf{E}\times \textbf{B}$ nonlinearity that couples different interacting modes. The modifications entailed in the present work will add to $\mathcal{L}$ the energy source for large-scale current-driven modes.

\subsection{Shifted Maxwellian Background Distribution}
Global tearing instabilities are driven by the current gradient in the plasma. However, the conventional \textsc{Gene} code assumes a Maxwellian as the background distribution, resulting in zero parallel current density under the first velocity moment. While more recent versions of the code offer non-Maxwellian distributions, e.g., the bi-Maxwellian \cite{di2018non}, they still yield a zero current density. To accurately model current-gradient-driven instabilities and replicate experimental observations, a shifted Maxwellian (SM) background distribution is employed to capture the realistic driving mechanism. Specifically, the background distribution for particle species $j$ is expressed as
\begin{equation}
F_{\mathrm{SM},j}=\frac{n_{0j}(x)\,m_j^{3/2}}{\left(2\pi T_{0j}(x)\right)^{3/2}}\exp\left[-\frac{(v_\parallel-v_{\parallel0,j}(x))^2}{2T_{0j}(x)/m_j}-\frac{\mu B_0(x)}{T_{0j}(x)}\right].
\label{shiftmax}
\end{equation}
In Eq.~(\ref{shiftmax}), $n_{0j}(x)$ and $T_{0j}(x)$ are the equilibrium density and temperature, $m_j$ is the particle mass, $B_0(x)$ is the equilibrium magnetic field, and $v_{\parallel0j}(x)$ is the velocity shift in the parallel direction, normalized to the thermal velocity $v_{\mathrm{Th},j}=[2T_{0j}(x_0)/m_j]^{1/2}$ at the reference location $x_0$ of the simulation domain. Although later not explicitly written, this parallel shift is $x$-dependent. Various terms in the linear operator of the gyrokinetic equation are modified according to the change in the background distribution function. One important term is the radial derivative of the background distribution, contributing to the gradient drive term,
\begin{eqnarray}
    \nabla_x F_{\mathrm{SM,}\,j} 
    =&-F_{\mathrm{SM,}\,j }\bigg\{\omega_{Tj}\left[\frac{(v_\parallel-v_{\parallel0,j})^2+\mu B_0}{2T_{0j}/m_j}-\frac{3}{2}\right]+\omega_{nj}\nonumber\\
    &-\frac{2(v_\parallel-v_{\parallel0,j})\nabla_x v_{\parallel0,j}-(v_{\parallel0,j}/v_\parallel)\mu\nabla_x B_0}{2T_{0j}/m_j}\bigg\}.
\end{eqnarray}
Here, $\omega_{Tj} \equiv -L_\mathrm{ref}\nabla_x\ln T_{0j}$ and $\omega_{nj} \equiv -L_\mathrm{ref}\nabla_x\ln n_{0j}$ are the temperature and density gradient scale lengths, respectively. These gradients are the energy source for well-known, particularly electrostatic, microinstabilities. The parallel velocity and the magnetic moment are denoted, respectively, $v_\parallel$ and $\mu$. We note that $L_\mathrm{ref}$ is a reference length, typically set to be the major radius $R_0$ for global calculations. The last term, involving the gradient of the shifted velocity, introduces the new driving force into the system, i.e., the current-gradient drive, as $\nabla v_{\parallel0}\propto \nabla J_{\parallel0}$.

\subsection{Modifications to the Field Equations}
The perturbed electrostatic potential $\phi_1(\textbf{x})$ can be evaluated using the Poisson equation
\begin{equation}
    \nabla_\perp^2\phi_1(\textbf{x}) = -4\pi\sum_jq_jn_{1,j}(\textbf{x}) = -4\pi\sum_jq_jM_{00,j}(\textbf{x}).
\end{equation}
Gaussian units are used, and all quantities are later normalized according to the \textsc{Gene} convention, with further details provided in Ref.~\cite{di2019implementation}. The perturbed density $n_{1,j}(\textbf{x})$ is expressed in terms of the zeroth moment $M_{00,j}(\textbf{x})$ of the distribution with respect to $v_\parallel$ and $\mu$. In gyrokinetics, the particle coordinates \textbf{x} can be decomposed into the gyration center \textbf{X} and the radial displacement  $\textbf{r}$ from the gyration center in the perpendicular plane, such that $\textbf{x}=\textbf{X}+\textbf{r}$. By employing the appropriate moments described in Refs.~\cite{di2018non, di2019implementation,Di_Siena_2018}, and neglecting parallel magnetic field fluctuations, the Poisson equation can be written as
\begin{eqnarray}
    \nabla_\perp^2\phi_1(\textbf{x})&=-\frac{8\pi^2qB_0}{m}\int\int\bigg\{\langle f_1(\textbf{x}-\textbf{r})\rangle\nonumber
    \\ 
    &+\left[\frac{\Omega}{B_0}\pder{F_\mathrm{SM}}{v_\parallel}-\frac{qv_\parallel}{cB_0}\pder{F_\mathrm{SM}}{\mu}\right]\left(A_{1,\parallel}(\textbf{x})-\langle\bar{A}_{1,\parallel}(\textbf{x}-\textbf{r})\rangle\right)
    \\
    &+\left[\frac{q}{B_0}\left(\phi_1(\textbf{x})-\langle\bar{\phi}_1(\textbf{x}-\textbf{r})\rangle\right)\right]\pder{F_\mathrm{SM}}{\mu}\bigg\}\dd v_\parallel\dd \mu.\nonumber
\end{eqnarray}
$\langle ... \rangle$ and overbars represent the gyroaverages defined as $\bar{\phi}_1=\langle\phi_1(\textbf{x})\rangle=\mathcal{G}[\phi_1(\textbf{x})]=\int \phi_1(\textbf{X}+\textbf{r}(\theta))\dd \theta/2\pi$, with the gyroradius vector $\textbf{r}(\theta)$ orthogonal to the magnetic field, and gyroangle $\theta$, $\langle \bar{\phi}_{1}(\textbf{x}-\textbf{r})\rangle$ is the double gyroaverage of $\phi_{1}$ defined as $\int\mathrm{d}\theta\int\mathrm{d}\textbf{X}\delta(\textbf{X}+\textbf{r}(\theta)-\textbf{x})\int\mathrm{d}\theta^\prime \phi_1(\textbf{X}+\textbf{r}(\theta^\prime))\mathrm{d}\theta^\prime/4\pi^2$, which in the global version is written as $\mathcal{G}^\dagger \phi_{1}(\textbf{x}-\textbf{r})\mathcal{G}$, as seen in some expressions later in the paper. This double gyroaverage arises from integration of the velocity moment $M_{00,j}$, which contains another gyroaverage of the perturbed distribution function \cite{di2019implementation}. The same applies to the electromagnetic vector potential $A_{1,\parallel}$ terms. Under the Coulomb gauge and in the absence of an equilibrium electric field, the parallel vector potential  is calculated from Amp\`ere's law and given by
\begin{equation}
 \nabla_\perp^2A_{1,\parallel}(\textbf{x}) =-\frac{4\pi}{c}\sum_j q_jv_{\parallel0,\,j}(\textbf{x}){n_{1,\,j}}(\textbf{x})= -\frac{4\pi}{c}\sum_j q_jM_{10,j}(\textbf{x}).
\end{equation}
This expression can be similarly written in terms of the shifted Maxwellian, yielding
\begin{eqnarray}
    \nabla_\perp^2A_{1,\parallel}(\textbf{x})&=-\frac{8\pi^2q}{mc}\int\int\bigg\{\langle f_1(\textbf{x}-\textbf{r})\rangle
    \nonumber\\ 
    &+\left[\Omega\pder{F_\mathrm{SM}}{v_\parallel}-\frac{qv_\parallel}{c}\pder{F_\mathrm{SM}}{\mu}\right]\left(A_{1,\parallel}(\textbf{x})-\langle\bar{A}_{1,\parallel}(\textbf{x}-\textbf{r})\rangle\right)
    \\
    &+q\left[\phi_1(\textbf{x})-\langle\bar{\phi}_1(\textbf{x}-\textbf{r})\rangle\right]\pder{F_\mathrm{SM}}{\mu}\bigg\}\dd v_\parallel \dd\mu.\nonumber
\end{eqnarray}
Combining Poisson's equation and Amp\`ere's law results in coupled field equations, which can be written in matrix form as 
\begin{equation}
\left[\matrix{ C_1 & C_2 \cr C_3 & C_4}\right]\left[\matrix{ \phi_1 \cr A_{1,\parallel}}\right]
=\left[\matrix{N_{00} \cr N_{10}}\right]
\label{coupled_equation}
\end{equation}
The potentials are solved by inverting the matrix $[C_i]$,
where
\begin{eqnarray*}
C_1  &= \frac{\pi q^2n_0}{T_0}\int\int \left[\frac{B_0F_\mathrm{SM}}{T_0(x)}\hat{\textbf{1}}-\mathcal{G}^\dagger\frac{B_0F_\mathrm{SM}}{T_0(x)}\mathcal{G}\right]\dd v_\parallel\dd \mu-\nabla_\perp^2\lambda_\mathrm{De}^2,
\\
C_2  &= \frac{2\pi q^2 n_0}{mv_\mathrm{Th}}\int\int\left[-\frac{B_0v_{\parallel 0}F_\mathrm{SM}}{T_0(x)}\hat{\textbf{1}}+\mathcal{G}^\dagger v_\parallel\frac{B_0F_\mathrm{SM}}{T_0(x)}\mathcal{G}\right]\dd v_\parallel\dd \mu,
\\
C_3  &=\frac{\pi q^2n_0\beta_\mathrm{ref}}{mv_\mathrm{Th}}\int\int v_\parallel\left[\frac{B_0F_\mathrm{SM}}{T_0(x)}\hat{\textbf{1}}-\mathcal{G}^\dagger \frac{B_0F_\mathrm{SM}}{T_0(x)}\mathcal{G}\right]\dd v_\parallel\dd \mu,
\\
C_4 &= -\nabla_\perp^2-\frac{\pi q^2n_0\beta_\mathrm{ref}}{m}\int\int \left[\frac{B_0v_\parallel v_{\parallel0}F_\mathrm{SM}}{T_0(x)}\hat{\textbf{1}}-\mathcal{G}^\dagger v_\parallel v_{\parallel0}\frac{B_0F_\mathrm{SM}}{T_0(x)}\mathcal{G}\right]\dd v_\parallel\dd \mu,
\\
N_{00}&=\pi q n_0 B_0\int\int \mathcal{G}(f_1(\textbf{x}))\dd v_\parallel \dd \mu,~\mathrm{and}
\\
N_{10}&=\pi qn_0\beta_\mathrm{ref}\frac{B_0v_\mathrm{Th}}{2}\int\int v_\parallel \mathcal{G}(f_1(\textbf{x}))\dd v_\parallel \dd \mu.
\end{eqnarray*}
The diagonal elements of the gyroaverage matrix are written as $\hat{\textbf{1}}$, while $T_0=T_0(x_0)$ and $n_0=n_0(x_0)$ are the temperature and density at the reference position $x_0$. The modified field equations couple the electrostatic and electromagnetic potentials, in contrast to the conventional equations derived from the Maxwellian distribution function, where the two are decoupled. This coupling allows for the consideration of the influence of a magnetic perturbation on the electrostatic field, as expected in self-consistent calculations of tearing modes \cite{furth1963finite, drake1977kinetic}. The re-derivation of velocity moments related to observable quantities, such as electromagnetic heat and particle fluxes, is also incorporated into the code to account for the modified transport caused by tearing instabilities.
\subsection{Parallel Current Profile}
The shifted Maxwellian distribution function requires as input the velocity shift profile $v_{\parallel0}(x)$, which can be evaluated from the equilibrium parallel current density based on the equation of the first velocity moment of the shifted Maxwellian
\begin{equation}
\textbf{J}_{\parallel0}(x)=\sum_j q_j \textbf{v}_{\parallel0,j}(x)n_{0j}(x)=\sum_j q_j\int \textbf{v}_{\parallel,j}F_{\mathrm{SM},j}\mathrm{d}^3\textbf{v}.
\label{totalCurrent}
\end{equation}
In the common scenario of a two-component-plasma current, ions and electrons tend to flow in opposite directions in response to an applied parallel electric field such that $\textbf{v}_{\parallel0,\mathrm{i}}=v_{\parallel0,\mathrm{i}}\hat{\textbf{b}}$ and $\textbf{v}_{\parallel0,\mathrm{e}}=v_{\parallel0,\mathrm{e}}(-\hat{\textbf{b}})$, with $\hat{\textbf{b}}$ the direction of the magnetic field. We assume that the flow velocities are on the order of the species' thermal velocities, with normalization $v_{\parallel0,j}=\hat{v}_{\parallel0}v_{\mathrm{Th,}j}$, where $\hat{v}_{\parallel0}$ a dimensionless quantity assigned to be equal for electrons and ions. This is equivalent to writing  $\textbf{v}_{\parallel0,\mathrm{i}}=\hat{v}_{\parallel0}v_\mathrm{Th,i}\hat{\textbf{b}}$ and $\textbf{v}_{\parallel0,\mathrm{e}}=-\hat{v}_{\parallel0}v_\mathrm{Th,e}\hat{\textbf{b}}$, implying $|v_{\parallel0,\mathrm{i}}/v_{\parallel0,\mathrm{e}}| = \sqrt{T_\mathrm{0i}m_\mathrm{e}/(T_\mathrm{0e}m_\mathrm{i})}$. In small-mass-ratio scenarios, the ion flow can be much smaller than the electron flow, but if not explicitly stated otherwise, all simulations performed in \textsc{Gene} employ the velocity-shift contribution from both species. 
Hence, the total current density can be written as 
\begin{equation}
    \textbf{J}_{\parallel0}(x)=\textbf{J}_{\parallel0,\mathrm{i}}(x)+\textbf{J}_{\parallel0,\mathrm{e}}(x)=q_\mathrm{i}\hat{v}_{\parallel0}v_\mathrm{Th,i}n_{0\mathrm{i}}\hat{\textbf{b}}-q_\mathrm{e}\hat{v}_{\parallel0}v_\mathrm{Th,e}n_{0\mathrm{e}}\hat{\textbf{b}}.
\end{equation}
Taking $q_\mathrm{i} = e$, $q_\mathrm{e} = -e$,  invoking the quasi-neutrality condition $n_{0\mathrm{e}}=n_{0\mathrm{i}}=n_0(x)$, the magnitude of total current density is written as 
\begin{eqnarray}
J_{\parallel 0}(x)&=en_{0}(x)\hat{v}_{\parallel0}(x)v_{\mathrm{Th,e}}\left[\sqrt{\frac{T_{0\mathrm{i}}}{T_{0\mathrm{e}}}\frac{m_\mathrm{e}}{m_\mathrm{i}}}+1\right],\label{jpar}
\end{eqnarray}
and thus the velocity shift reads
\begin{eqnarray}
\hat{v}_{\parallel0}(x) &= \frac{J_{\parallel0}(x)}{en_{0}(x)v_{\mathrm{Th,e}}}\left[\sqrt{\frac{T_{0\mathrm{i}}}{T_{0\mathrm{e}}}\frac{m_\mathrm{e}}{m_\mathrm{i}}}+1\right]^{-1}.\label{vshifted}
\end{eqnarray}
The current can also be extracted from the equilibrium profile of the safety factor $q$, as the $q$-profile contains information about the toroidal and poloidal magnetic fields. 
To enable benchmarks for established tearing mode behavior in tokamak geometry, we develop expressions relating $q$ to the current that is appropriate for a large-aspect-ratio tokamak. Later, when the RFP is considered, we will directly use experimentally relevant current profiles generated from the equilibrium solver MSTfit.
For a large-aspect-ratio circular cross-section, the $q$-profile can be written in terms of the cylindrical approximation as
\begin{equation}
    q(r) = \frac{r}{R_0}\frac{B_\varphi(r)}{B_\mathrm{\theta}(r)},
\end{equation}
where $B_\varphi(r)$ and $B_\theta(r)$ are toroidal and poloidal magnetic fields, respectively, and $r$ denotes the radial coordinate, which from this point can be used interchangeably with $x$. Starting with Amp\`{e}re's law $(4\pi/c)\textbf{J}=\nabla\times\textbf{B}$, the cylindrical approximation with $B_\theta\ll B_\varphi\sim B_0$ yields a parallel current density given by
\begin{eqnarray}
J_{\parallel0}(r) \approx \frac{c}{4\pi}\left[\frac{1}{r}\pder{(rB_\theta)}{r}\right] = \frac{cB_0}{4\pi R_0}\frac{\partial }{r\partial r}\left[\frac{r^2}{q(r)}\right].
\end{eqnarray}
The prefactor $cB_0/(4\pi R_0)$ is equivalent to the current density $J_{\parallel0}$ at the axis. Equating this expression to Eq.~(\ref{vshifted}), the velocity shift can be written as 
\begin{equation}
    \hat{v}_\mathrm{\parallel0}(r) = \frac{c B_0/(4\pi R_0)}{n_{0}(r)ev_{\mathrm{Th,e}}}\left[\sqrt{\frac{T_{0\mathrm{i}}}{T_{0\mathrm{e}}}\frac{m_\mathrm{e}}{m_\mathrm{i}}}+1\right]^{-1}\left[\frac{\partial }{r\partial r}\left(\frac{r^2}{q}\right)\right].
\end{equation}
 In terms of normalized parameters, the dimensionless velocity shift is equivalently given by
\begin{equation}
\hat{v}_{\parallel 0}(r)    =\frac{\sqrt{2}\varepsilon_{\mathrm{a}}\rho_\mathrm{i}^*}{\hat{n}_{0}(r)\beta_\mathrm{e}}\left[1+\sqrt{\frac{T_{0{\mathrm{e}}}}{T_{0{\mathrm{i}}}}\frac{m_\mathrm{i}}{m_\mathrm{e}}}\right]^{-1}\left[\frac{\partial }{r\partial r}\left(\frac{r^2}{q(r)}\right)\right],
    \label{vshift_q}
\end{equation}
where $\beta_\mathrm{e}=8\pi n_{0\mathrm{e}}T_{0\mathrm{e}}/B_0^2$, $\rho_\mathrm{i}^*=\rho_\mathrm{i}/a$, $\varepsilon_a=a/R_0$, and $\hat{n}_0(r) = n_0(r)/n_0(r_0)$, with $r_0$ the reference position in the simulation domain. Also note that the electron skin depth $\delta_\mathrm{e}$ can be written directly as
\begin{equation}
\frac{\delta_\mathrm{e}}{a} =\sqrt{\frac{m_\mathrm{e}}{m_\mathrm{i}}\frac{T_{0\mathrm{e}}}{T_{0\mathrm{i}}}}\frac{\rho_{\mathrm{i}}^*}{\sqrt{\beta_\mathrm{e}/2}},
\label{esd}
\end{equation}
and that this form relates to the strength of velocity shift as shown in Eq.~(\ref{vshift_q}). The electron skin depth is related to the singular layer width for collisionless tearing modes, allowing the determination of tearing mode behavior and stability regimes \cite{drake1977kinetic, porcelli1991collisionless}.
\section{Benchmarking of the Current-Gradient Drive}
The implementation of a shifted Maxwellian based on a velocity shift in the Vlasov-Maxwellian system of equations represents a significant modification to \textsc{Gene}, enabling self-consistent global simulations of current-driven instabilities. Before deploying this upgraded code to study linear tearing instability, nonlinear tearing evolution, and later multi-scale interactions in the RFP, it is important to test the fidelity of the implementation against different computational models. The analytic forms of the equilibrium profiles and other input parameters, for which tearing modes are unstable, are listed in Table~\ref{table_parameters}.
\begin{table}
  \centering
  \scriptsize{
  \begin{tabular}{l | c | c | c }
  \hline\hline
    \multirow{2}{*}{Equilibrium} &  \multirow{2}{*}{ORB5} & GKW & Two-Fluid \\
    & & Hornsby \textit{et al}. \cite{hornsby2015linear,hornsby2015non} &Ishizawa \textit{et al}. \cite{ishizawa2009thermal}\\\hline\hline & & & \\[-1.2ex]
    $q(x_a)$ & $\sum_{i=0}^8c_ix_a^i$ & $3.5x_a^2[1-(1-x_a^2)^2]^{-1}$ & $1.7(1+29.65x_a^8)^{1/4}$\\
    $T_\mathrm{0i}=T_\mathrm{0e}=T_\mathrm{eq}(x_a)$ & 1 & $(R/L_T)\int [\tanh(100x_a-5)-\tanh(100x_a-105)]/2\dd x_a$& $0.8-0.2\tanh(5x_a-3.5)$\\
    $n_\mathrm{0i}=n_\mathrm{0e}=n_\mathrm{eq}(x_a)$ & 1 & $(R/L_n)\int [\tanh(100x_a-5)-\tanh(100x_a-105)]/2\dd x_a$ &$0.8+0.2\exp(-4x_a^2)$\\
    $\beta_\mathrm{e}=8\pi n_\mathrm{0e}T_\mathrm{0e}/B_0^2$ 
                    & 0.2\% & 0.1\% &1\%\\
    $\rho_\mathrm{i}^*=\rho_\mathrm{i}/a$ & $0.01$ & $0.0118$ & $0.0125$\\
    $\varepsilon_a=a/R_0$ & $0.1$ & $0.3$ & $0.25$\\
    $m_\mathrm{e}/m_\mathrm{i}$ & $0.005$ & $0.000544$ & $0.000544$\\
    $\nu_\mathrm{c}$ & 0 & 0.00112 &0.03187\\[1.2ex]
   \hline && \\[-1.2ex]
   domain~[$x_a^\mathrm{min}$, $x_a^\mathrm{max}$] & [0.01, 0.99] & [0.1, 1]&[0.01, 0.99]\\
   buffers [left, right] & [0.1$a$,\,0.1$a$]& [0.05$a$,\,0.05$a$] & [0.05$a$,\,0.05$a$]\\
$D_x$,\,$D_z$,\,$D_v$&$0.2,\,0.2,\,0.2$& $0.1,\,3.4,\,0.2$ & $0.05,\,3.4,\,0.1$\\
%    \verb!lv! & 3.0 & 3.0 & 3.0\\
%    \verb!lw! & 6.0 & 9.0 & 9.0\\
%    \verb!nx0! & 512 & 512 & 256\\
%    \verb!nz0! & 16 & 16& 16\\
%    \verb!nv0! & 32 & 32& 16\\
%    \verb!nw0! & 8 & 9& 16\\
%    \verb!x0! & 0.5 & 0.55& 0.5\\
    [1.2ex]
    \hline\hline
  \end{tabular}
  \caption{\label{table_parameters} Analytic forms of equilibrium profiles and parameters used for benchmarks in this work. Note that $x_a=x/a$ is the normalized radius. $q$, $T_\mathrm{eq}$, $n_\mathrm{eq}$, $\beta_\mathrm{e}$, $\rho_\mathrm{i}$, $a$, $R_0$, $m_\mathrm{e}$, and $m_\mathrm{i}$ are safety factor, equilibrium temperature, equilibrium density, plasma beta, ion gyroradius, device minor radius, major radius, electron mass, and ion mass, respectively. Note for the safety factor profile in the ORB5 case, $q(x_a)=\sum_{i=0}^8 c_ix_a^i$, with coefficients $c_i$$=$($1.4950$, $-0.0052$, $0.7969$, $-1.2433$, $6.0576$, $-14.4208$, $20.4868$, $-15.1526$, $4.8045$). The bottom half panel displays the boundary constraints \cite{gorler2011global} and hyperdiffusivities \cite{pueschel2010dissipation} set in \textsc{Gene}. $x_a^\mathrm{min}$ and $x_a^\mathrm{max}$ are the left and the right ends of the simulation domain, respectively, \textit{buffer} refers to the size of the left and right buffer zones, measured in minor radius $a$, and $D_x$, $D_z$, and $D_v$ are the hyperdiffusivities applied to the $x$, $z$, and $v_\parallel$ discretization, respectively.}}
\end{table}
\subsection{Comparisons with ORB5}
Linear global tearing modes simulated in the \textsc{Gene} code are compared with the results obtained from the gyrokinetic PIC code ORB5. Prior to this investigation, both codes have demonstrated successful benchmarking for scenarios involving global ion temperature gradient (ITG) and kinetic ballooning modes (KBMs) \cite{gorler2016intercode}, geodesic acoustic modes (GAMs) \cite{biancalani2017cross}, energetic-particle-driven GAMs (EGAMs) \cite{Di_Siena_2018}, and toroidal Alfv\'en eigenmodes (TAEs) \cite{di2021nonlinear} which later was extended to the Alfv\'en-ITG system \cite{disiena2023how}. Notably, a recent extension of the ORB5 code incorporates the current-gradient drive \cite{mishchenko2019pullback}, similar to the capabilities of \textsc{Gene} presented here. Consequently, this benchmarking exercise presents the first simulations of linear tearing modes using the ORB5 code.

\begin{figure}[ht!]
    \centering
    \includegraphics[scale=0.43]{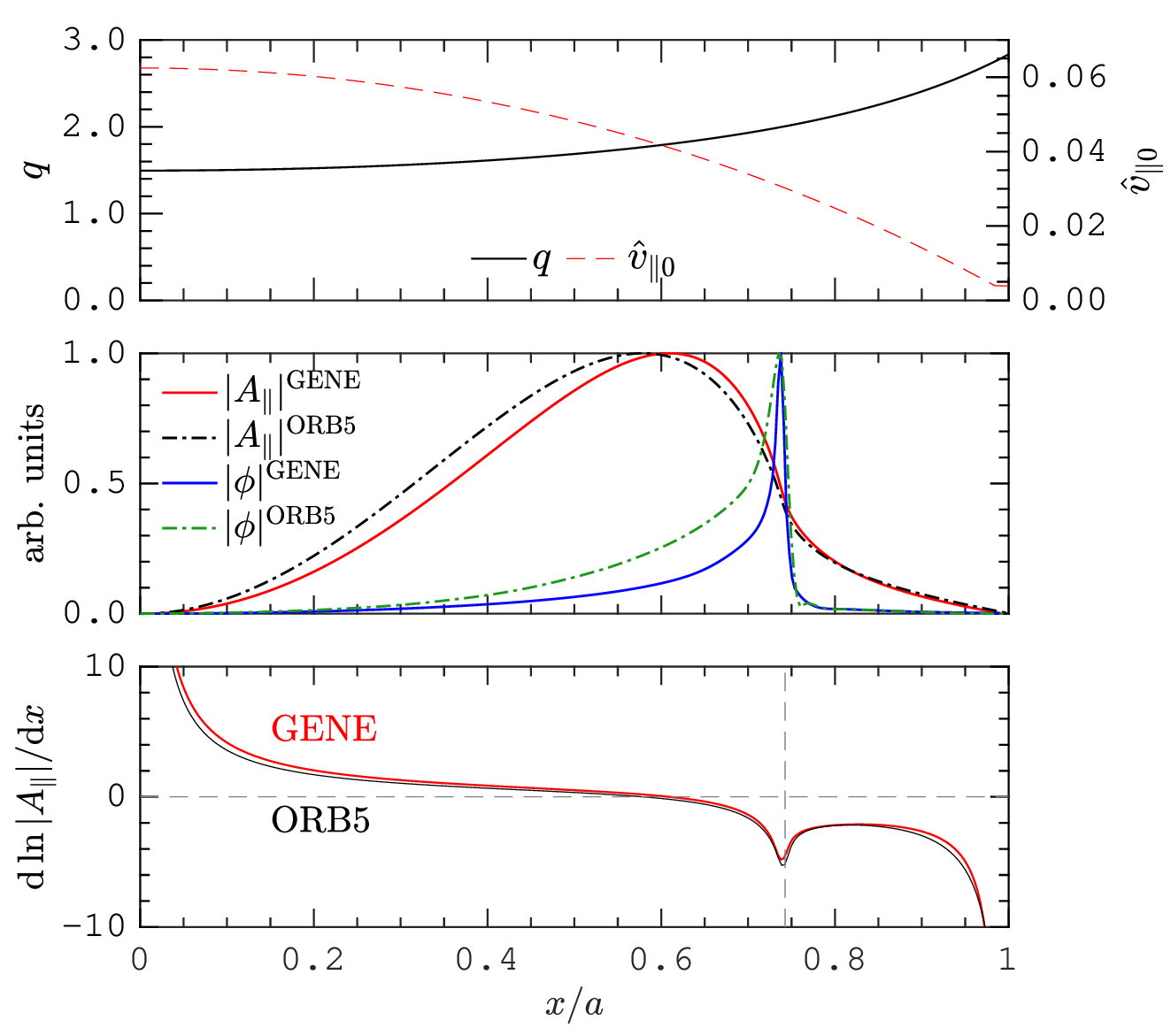}
    \caption{[Top] Profiles of the safety factor (black solid line, left axis) and the velocity shift (red dashed line, right axis) used for the benchmark with ORB5, with plasma parameters given in Table~\ref{table_parameters}. [Middle] The normalized radial eigenmode structures simulated in \textsc{Gene} are presented in comparison with ORB5, showing tearing features at the rational surface. [Bottom] Comparison of the normalized radial gradient of $A_\parallel$ between the two codes, showing an abrupt change in the structure of $A_\parallel$, confirming a $(m,\,n)=(1,\,2)$ tearing mode centered at $x/a\approx0.74$.}
    \label{ORB5EquilibriumProfile}
\end{figure}

The equilibrium safety factor $q(x_a)$, where $x_a=x/a$ and $a$ is a minor radius, for this test is an 8th-order polynomial which is unstable to an $m/n=2/1$ tearing mode, where $m$ and $n$ are the poloidal and toroidal mode numbers, respectively. The polynomial coefficients $c_i$ are given in Table~\ref{table_parameters}, and the corresponding profiles are shown in Fig.~\ref{ORB5EquilibriumProfile}, including the velocity shift profile calculated  under the cylindrical approximation, see Eq.~(\ref{vshift_q}). While simulations in \textsc{Gene} include velocity shifts for electrons and ions, ORB5 considers only the contribution from electrons; this, however, results in only minute differences, less than the convergence threshold for either code. The background temperature and density are assumed to be radially constant to avoid pressure-gradient-driven modes and diamagnetic effects. A small-inverse-aspect-ratio concentric circular geometry is used, making the cylindrical approximation justifiable and minimizing curvature effects. This makes comparison of the numerical growth rates with theoretical scalings of slab current sheets possible \cite{drake1977kinetic,porcelli1991collisionless}. For simplicity, the standard case sets collisionality $\nu_\mathrm{ei}$ to zero, resulting in a collisionless tearing mode. A list of other plasma parameters is given in Table~\ref{table_parameters}.
\begin{figure}
    \centering
    \includegraphics[scale=0.45]{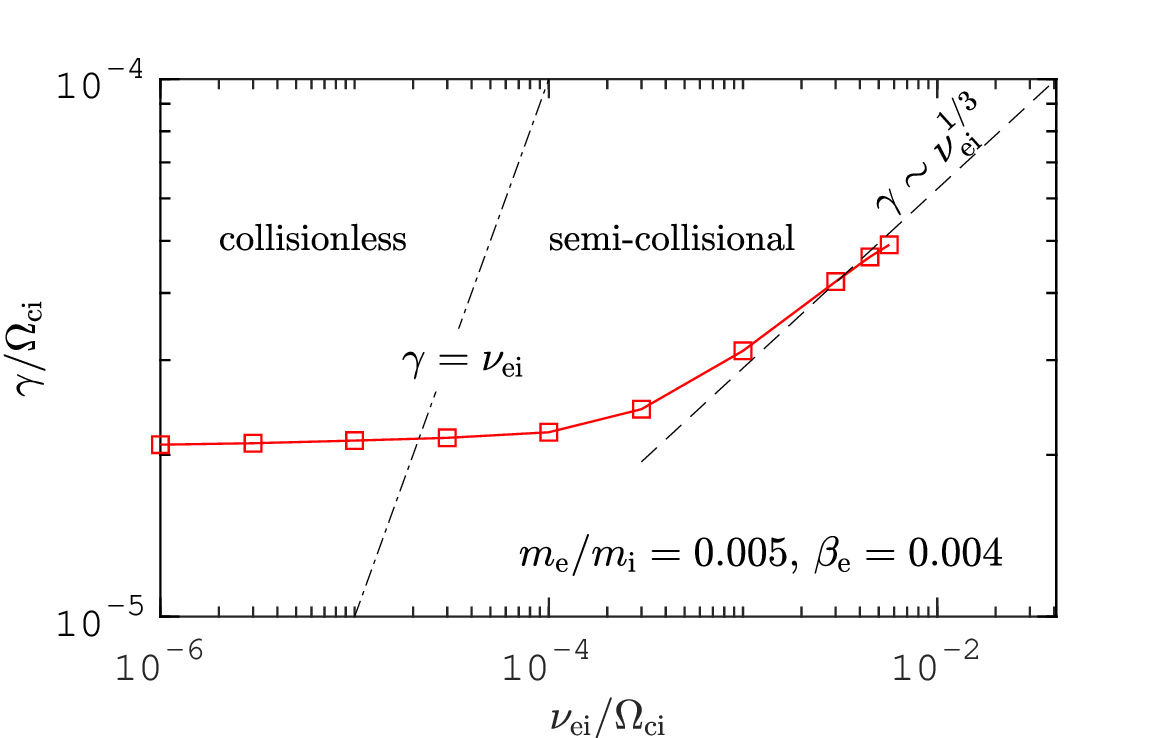}
    \caption{Tearing-mode growth rate as a function of electron-ion collision frequency, computed with \textsc{Gene}. Scaling consistent with theoretical prediction is shown; i.e., no dependence in the collisionless regime and $\gamma\propto \nu_\mathrm{ei}^{1/3}$ in the semi-collisional regime.}
    \label{ORB5:collisional-scan}
\end{figure}
\begin{figure}[ht!]
    \centering
    \includegraphics[scale=0.45]{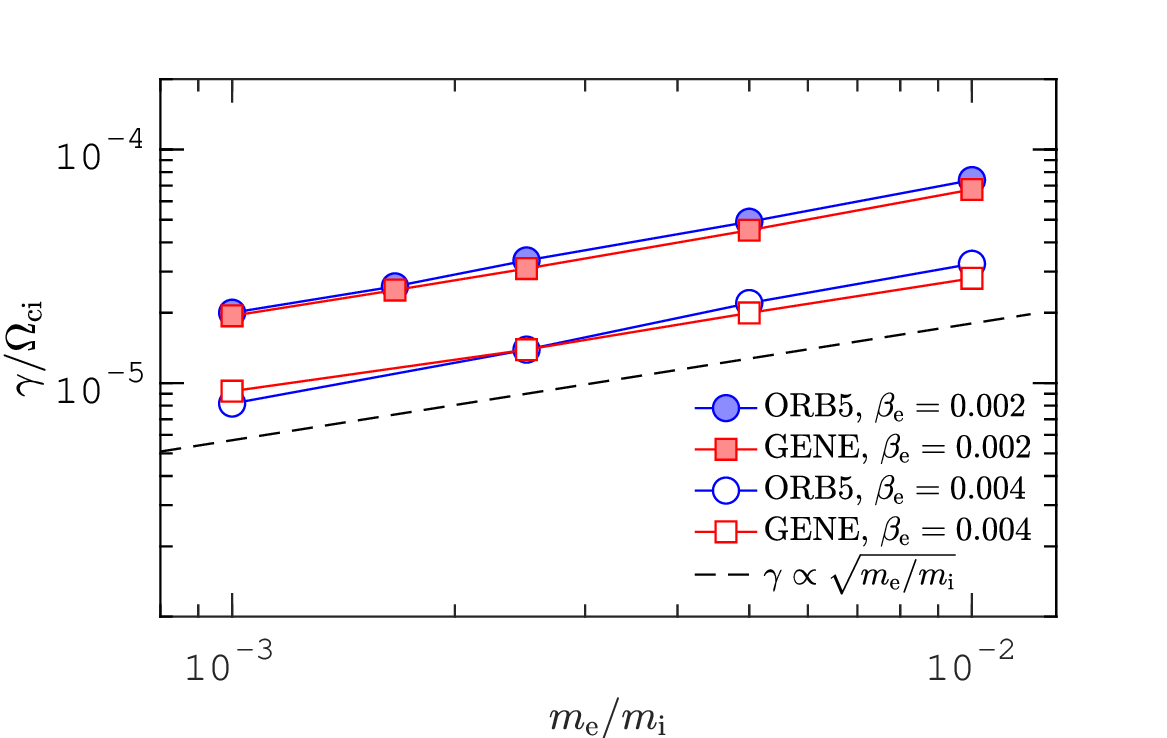}
    \caption{Mass ratio scan, simulations from \textsc{Gene} (red squares) and ORB5 (blue circles) at two different vaules of $\beta_\mathrm{e}=\beta_{\mathrm{e}}^{\textsc{Gene}}= 8\pi n_\mathrm{e} T_\mathrm{e}/B_0^2$. Note that  $\beta_{\mathrm{e}}^{\textsc{Gene}}=2\beta_{\mathrm{e}}^{\mathrm{ORB5}}$. The theoretical growth rate scaling $\gamma\propto \sqrt{m_\mathrm{e}/m_\mathrm{i}}$ is shown as the black dashed line. Good agreement is observed between codes and theory.}
    \label{ORB5:mass-scan}
    \includegraphics[scale=0.45]{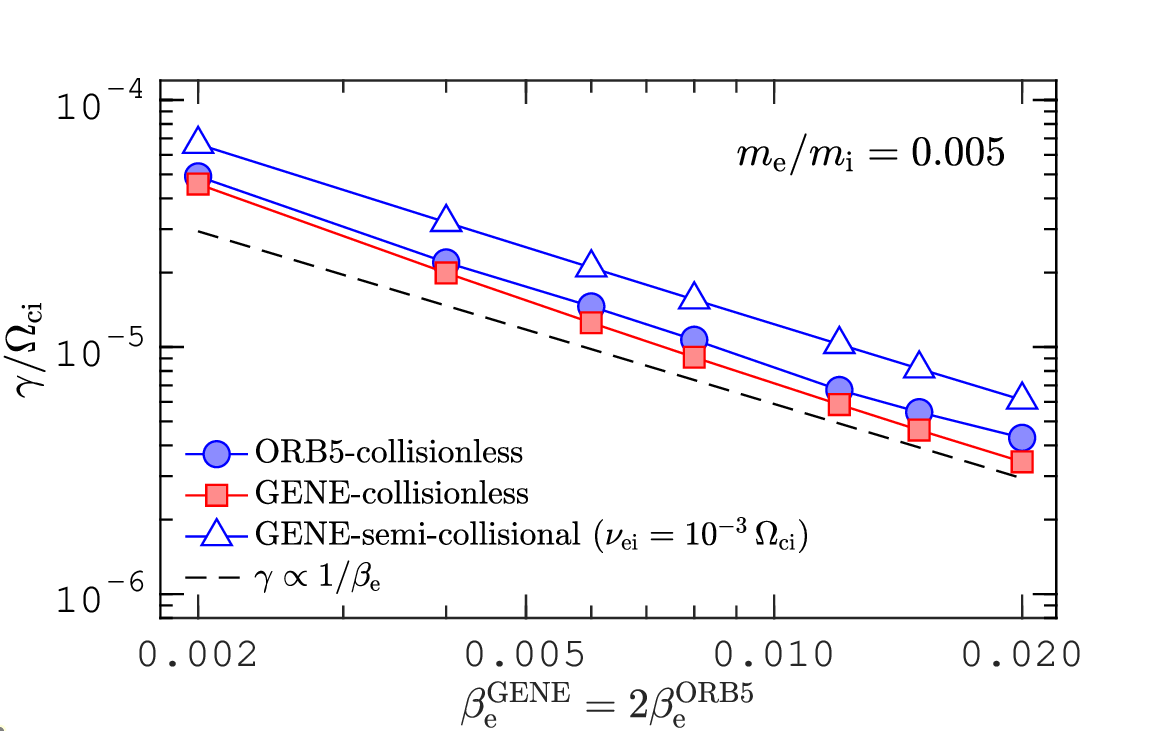}
    \caption{Collisionless $\beta_\mathrm{e}$ scans performed with \textsc{Gene} (red squares) and ORB5 (blue circles). The two codes agree with each other and with the theoretical scaling $\gamma\propto 1/\beta_\mathrm{e}.$ Blue triangles show \textsc{Gene} data in the semi-collisional regime where the same scaling is recovered.}
    \label{ORB5:beta-scan}
\end{figure}

For $n=1$, corresponding to $k_y\rho_\mathrm{s}=0.034$, \textsc{Gene} produces a tearing mode  with growth rate $\gamma_\textsc{Gene}=0.045\,c_\mathrm{s}/R_0$ = $4.5\times 10^{-5}\Omega_\mathrm{ci}$, where $c_\mathrm{s}=\sqrt{T_{0\mathrm{e}}/m_\mathrm{i}}$ is ion acoustic speed, and $\Omega_\mathrm{ci}=(c_\mathrm{s}/R_0)/(\varepsilon_a\rho_\mathrm{i}^*)$ is the ion cyclotron frequency. ORB5 gives $\gamma_\mathrm{ORB5}=4.9\times10^{-5}\,\Omega_\mathrm{ci}$, in good agreement with \textsc{Gene}. The growth rates are very small compared to the cyclotron frequency, and are roughly an order of magnitude lower than those of ion-scale instabilities, correctly reflecting the characteristic time scale of tearing-mode fluctuations. The radial $\phi$ and $A_\parallel$ eigenmode structures from \textsc{Gene} are shown in the middle panel of Fig.~\ref{ORB5EquilibriumProfile} and trace with those of ORB5. Discontinuities across the rational surface $q=2$ of $\phi$ and $A_\parallel$ showing $\Delta^\prime>0$ are well-known indications of an unstable tearing mode.

In MHD, tearing modes are destabilized by resistivity \cite{hornsby2015linear, furth1963finite,drake1977kinetic,porcelli1991collisionless}. Here, the collision frequency, which relates to resistivity in MHD, is varied, and the response of the tearing mode is recorded. Collisionality in \textsc{Gene} for this paper, except where stated otherwise, is modeled by a Landau collision operator to reduce the computational cost; other collision operators produce comparable growth rates but with higher computational expense. The normalized collision frequency used in \textsc{Gene} is defined as 
\begin{equation}
    \nu_\mathrm{coll} = \frac{\pi \ln \Lambda\, e^4 n_\mathrm{ref} L_\mathrm{ref}}{2^{3/2}T_\mathrm{ref}^{2}},
\end{equation}
where constituting quantities are given in Gaussian units. $\ln \Lambda$ is the Coulomb logarithm, which can be expressed as $24-\ln\left(\sqrt{ 10^{13}n_\mathrm{ref}}/(10^{-3}T_{\mathrm{ref}})\right)$, with the reference value of background density $n_\mathrm{ref}$ in units of $10^{19}\mathrm{m}^{-3}$ and temperature $T_{\mathrm{ref}}$ in units of keV, and $L_\mathrm{ref}$ is the reference length in meters. The electron-ion collision frequency can be written in units of $c_\mathrm{s}/R_0$ as 
\begin{equation}
   \nu_\mathrm{ei}=4Z^2\sqrt{\frac{m_\mathrm{i}}{m_\mathrm{e}}}\frac{c_\mathrm{s}}{R_0}\nu_\mathrm{coll},
 \end{equation}
with the ion charge number $Z$ and sound speed $c_\mathrm{s}=\sqrt{T_\mathrm{0e}/m_\mathrm{i}}$, and $m_\mathrm{i}$ and $m_\mathrm{e}$ are ion and electron mass, respectively. $\nu_\mathrm{ei}$, hence, has the same unit as growth rate. The growth rate of the tearing mode as a function of collisionality is plotted in Fig.~\ref{ORB5:collisional-scan}. Two collisionalality regimes are evident, a collisionless and a semi-collisional regime. In the limit of collisionless tearing, growth rates approach a constant when $\nu_\mathrm{ei}\ll\gamma$, while in the semi-collisional regime $\nu_\mathrm{ei}\sim\gamma$, growth rates approach a scaling of $\gamma\propto \nu_\mathrm{ei}^{1/3}$ \cite{drake1977kinetic, porcelli1991collisionless}. Simulations at higher collision frequencies were also attempted to see if the collisional regime could be reached; however, numerical instability prevented further study of this regime. This scaling behavior confirms that the instability reproduced by the modified \textsc{Gene} is indeed a tearing mode.

To assess the agreement between \textsc{Gene} and ORB5 across varying plasma parameters and to investigate the scaling behavior of tearing modes in accordance with theoretical predictions, systematic scans of mass ratios at different $\beta_\mathrm{e}$ are conducted using the initial setup, focusing on the standard collisionless case. Fig.~\ref{ORB5:mass-scan} presents the tearing-mode growth rates at different mass ratios, compared with the corresponding computations from ORB5. Not only do the growth rates obtained from the two codes exhibit good agreement, but they also exhibit consistent scaling with the theoretical prediction $\gamma\propto \sqrt{m_\mathrm{e}/m_\mathrm{i}}$ \cite{drake1977kinetic}. Additionally, a $\beta$ scan is performed and compared with the results from ORB5 as depicted in Fig.~\ref{ORB5:beta-scan}, which shows that growth rates follow each other and fall into correct theoretical scaling $\gamma\propto1/\beta_\mathrm{e}$ \cite{drake1977kinetic}. Also confirmed by this figure is the fact that \textsc{Gene} is capable of reproducing the expected $\beta$-scaling in the semi-collisional limit.

\subsection{Comparison with GKW}

Simulations of tearing modes are conducted to compare the results with published findings from another gyrokinetic code, GKW \cite{peeters2009nonlinear}. Similar to \textsc{Gene}, GKW is a global gyrokinetic code that uses a $\delta f$-approach capable of computing current-gradient-driven instabilities \cite{ho1991nonlinear}. Previous studies employing GKW have examined the linear and nonlinear evolution of global-scale tearing modes in tokamak systems using a shifted-Maxwellian distribution function \cite{hornsby2015seed,hornsby2015linear,hornsby2015non}. Detailed information regarding code implementation can be found in those publications, noting the distinction that the field equations are coupled in \textsc{Gene} but decoupled in GKW. The decoupling arises from a very small magnitude of the velocity shift $v_{\parallel 0}\ll 1$, causing coefficients $C_2$ and $C_3$ in Eq.~(\ref{coupled_equation}) to become negligible. The approximation breaks down when the shifted velocity becomes significant relative to the thermal velocity. The assumption retains its validity within the scope of this benchmark due to the smallness of velocity shift; notwithstanding, a notable velocity gradient, or equivalently current gradient, exists. The successful simulation of tearing-mode evolution in GKW establishes it as a suitable computational tool for comparison with \textsc{Gene}.

\begin{figure}[ht!]
    \centering
 \includegraphics[scale=0.46]{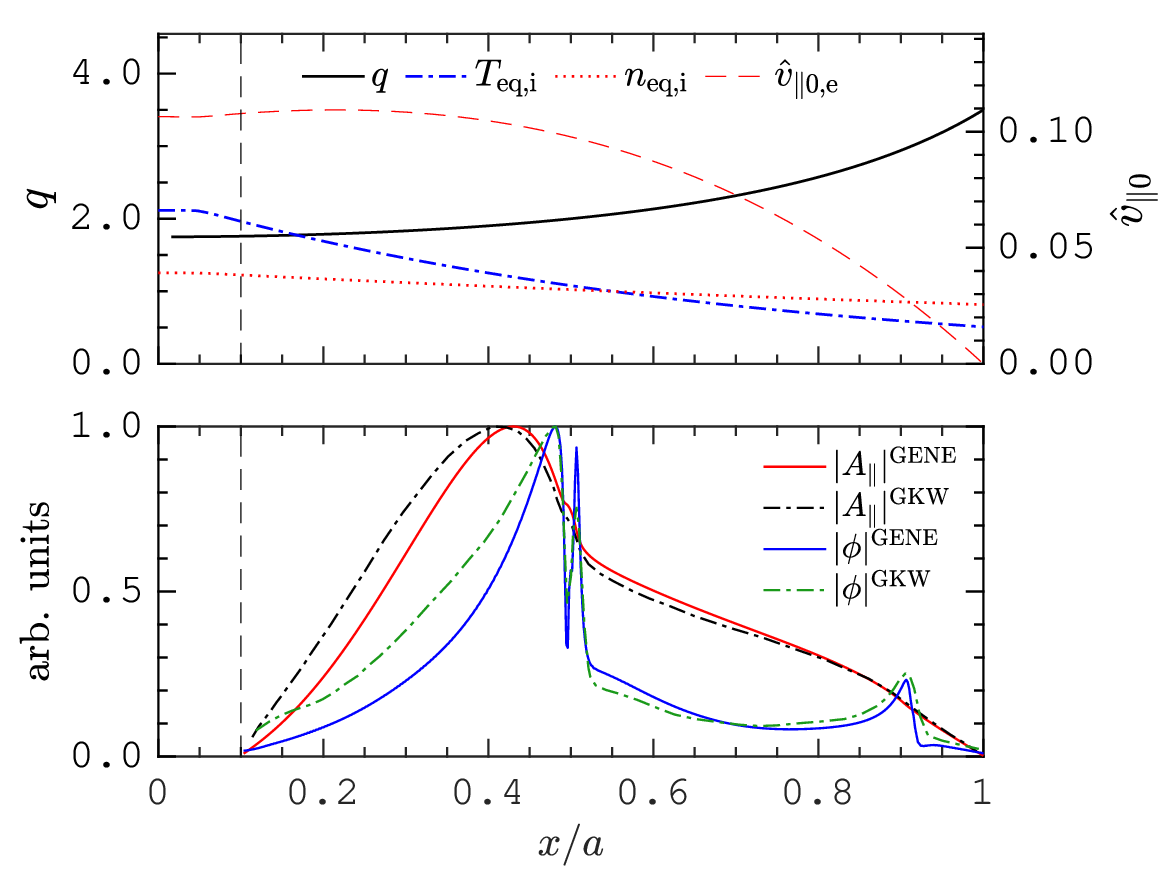}
     \caption{[Top] Equilibrium quantities used for the benchmark with GKW \cite{hornsby2015linear,hornsby2015non} (case 3 in Table 2). Profiles of safety factor (black solid line), temperature (blue dash-dotted), and density (red dotted) are referenced on the left axis, and velocity shift (red dashed) is referenced on the right axis. [Bottom] The corresponding normalized radial eigenmode structures obtained from \textsc{Gene} in comparison with GKW.}
    \label{fig:HBprofiles}
\end{figure}

The present benchmark is based on the equilibrium and parameters utilized by Hornsby \textit{et al.} \cite{hornsby2015linear,hornsby2015non}, as also listed in Table~\ref{table_parameters}. A concentric circular magnetic geometry is adopted, with temperature and density profiles, as shown in the top panel of Fig.~\ref{fig:HBprofiles}. It is worth noting that the normalization differs between the two codes, such that $\rho^*_{\textsc{Gene}}=(\rho^*_{\mathrm{GKW}}/\sqrt{2})(R/a)$. Moreover, it is important to highlight that the study by Hornsby \textit{et al.} solely focuses on an electron current density and does not consider an ion current density. Hence we apply the same convention in \textsc{Gene} for this particular benchmark. 

The radial tearing-mode structures obtained from \textsc{Gene}, depicted in the bottom panel of Fig.~\ref{fig:HBprofiles}, match those from GKW. These mode structures exhibit a fundamental tearing feature, characterized by abrupt changes in radial mode structures at rational surfaces, i.e., $q=2/1$ at $x/a=0.5$ and $q=3/1$ at $x/a=0.9$. The ratio of the peaks of electrostatic to electromagnetic potentials, as determined by \textsc{Gene}, is comparable to the ratio reported in Ref.~\cite{hornsby2015non} for GKW. A comparison of growth rates and frequencies at different values of plasma $\beta$ and pressure gradients, as presented in Table 2, demonstrates good agreement between the two codes. Slight deviations in Cases 4 and 5 can be attributed to small growth rates, as they fall within the range of computational error bars. 

The comparison shows that GKW and \textsc{Gene}, though implemented slightly differently, agree on global-tearing growth rates and frequencies, even under the influence of diamagnetic effects, i.e., density and temperature gradients. The growth rate dependence on $R/L_T$ also follows the analytical predictions made by Ref.~\cite{cowley1986linear} and a unified theory of internal kink and tearing mode instabilities \cite{connor2012unified}.
\begin{table*}[ht]
  \centering
  \footnotesize{
  \begin{tabular}{c|c|c|c|c|c|c|c}
  \hline \hline
  \multirow{2}{*}{Case} & \multirow{2}{*}{~$\beta_\mathrm{e}(\%)$~} & \multirow{2}{*}{~$R/L_n$~} & \multirow{2}{*}{~$R/L_T$~} & \multicolumn{2}{c|}{GKW}& \multicolumn{2}{c}{\textsc{Gene}} \\\cline{5-8}
  & & & & ~$\gamma/[v_\mathrm{Th,i}/R_0]$~ & ~$\omega/[v_\mathrm{Th,i}/R_0]$ ~& ~$\gamma/[v_\mathrm{Th,i}/R_0]$~ & ~$\omega/[v_\mathrm{Th,i}/R_0]$~ \\
  \hline\hline & & & & & & & \\[-1.5ex]
  1 & 0.05 & 1.5 & 5.0 & $0.045$& $-0.16$ &$0.048$ & $-0.16$\\
  2 & 0.10 & 1.0 & 3.5 & $0.032$& $-0.12$ &$0.031$ &$-0.11$ \\
  3 & 0.10 & 1.5 & 5.0 & $0.024$& $-0.15$ &$0.023$ & $-0.17$\\
  4 & 0.20 & 1.5 & 5.0 & $0.009$& $-0.15$ &$0.012$ & $-0.09$\\
  5 & 0.10 & 2.2 & 6.9 & $0.0032$& $-0.28$ &$0.0070$ & $-0.24$\\[-1.5ex]
  & & & & & & & \\\hline\hline
  \end{tabular}}
  \caption{Comparison of growth rates and frequencies between GKW and \textsc{Gene} at different plasma $\beta$s and gradients. Here, $R/L_n$ and $R/L_T$ are density and temperature gradient scale lengths, and $v_\mathrm{Th,i}=\sqrt{2T_\mathrm{i}/m_\mathrm{i}}$}
  \label{table_comparison}
\end{table*}
\subsection{Comparison with Fluid Model}

A resistive reduced two-fluid model has successfully described tearing modes and has been utilized to investigate the interactions between large-scale tearing modes and small-scale KBM \cite{ishizawa2009thermal}. It is useful to compare tearing modes modeled with the gyrokinetic model with those produced by the fluid model. Hence, we adopt the safety factor and pressure profiles employed in Ref.~\cite{ishizawa2009thermal} in this benchmark. These equilibrium profiles have been previously demonstrated to generate tearing modes for the investigation of thermal transport in the presence of magnetic fluctuations, as well as island formation and rotation \cite{ishizawa2019multi}. 

\begin{figure}[ht!]
    \centering
    \includegraphics[scale=0.46]{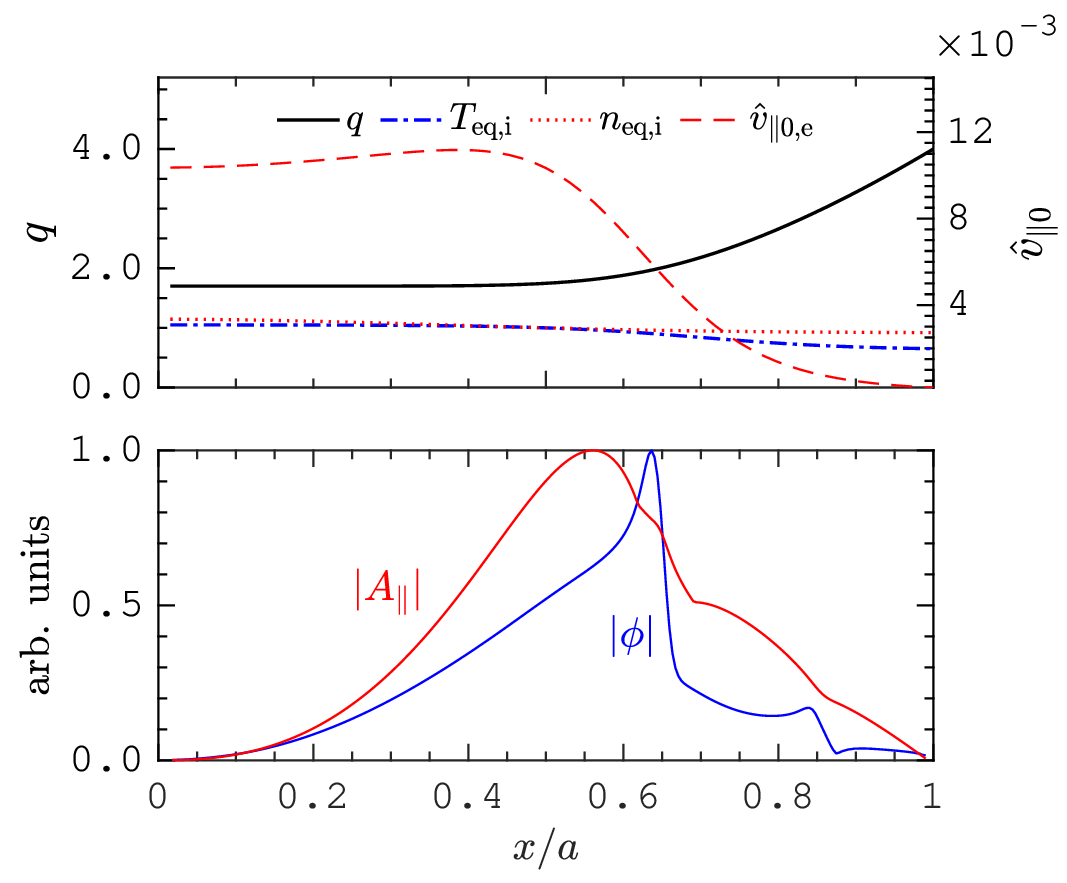}
    \caption{[Top] Equilibrium quantities used for the benchmark with a two-fluid model \cite{ishizawa2009thermal}. The left axis are profiles of safety factor (black solid), temperature (blue dash-dotted), and density (red dotted); on the right axis is the velocity shift (red dashed). [Bottom] The corresponding normalized radial eigenmode structures of the $n=1$ tearing mode}
    \label{fig:ishizawaprof}
\end{figure}

Table 1 shows the analytical expressions of the safety factor, temperature, and density profiles. The safety factor at the magnetic axis is $q_0=1.7$, and the value at the minor radius is $q_a=4.0$, as illustrated in the top panel of Fig.~\ref{fig:ishizawaprof}. The corresponding shifted velocity calculated from Eq.~(\ref{vshift_q}) is also shown. Notably, the $q$-profile exhibits resonance of the tearing mode at the $q=2$ surface located at $x/a = 0.65$.

\begin{figure}[ht!]
    \centering
    \includegraphics[scale=0.35]{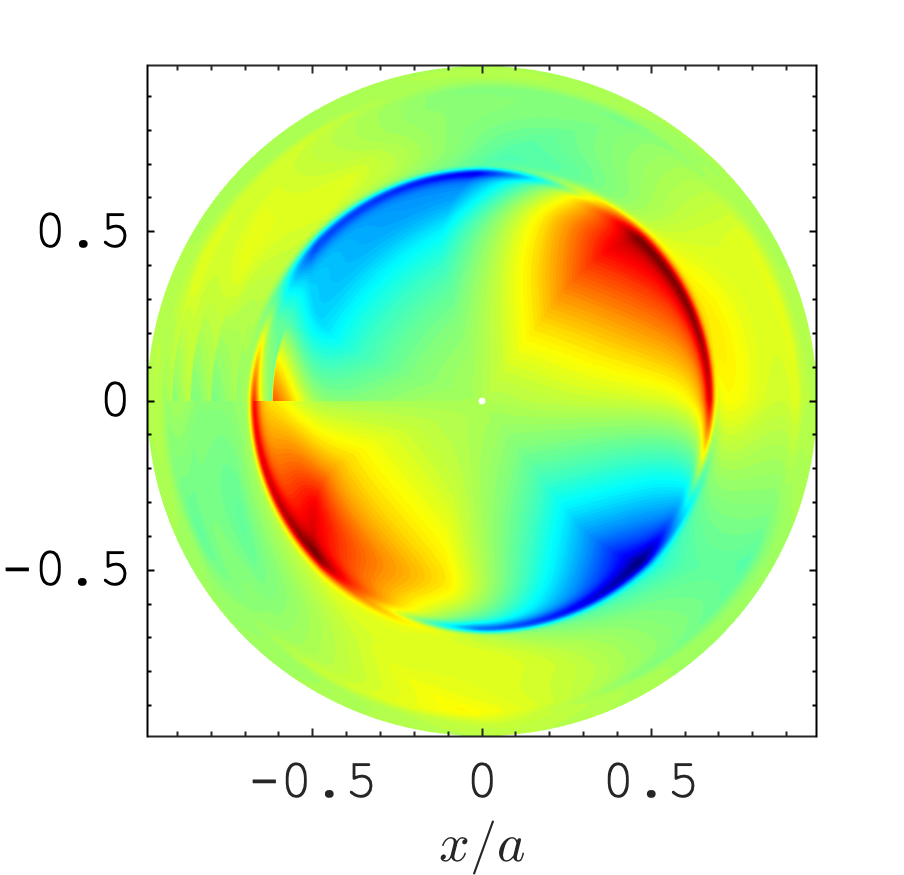}
    \caption{A poloidal cross-section of the electrostatic potential of the $n=1,~m=2$ tearing mode obtained from \textsc{Gene} for the equilibrium used in the benchmark against the two-fluid model. This poloidal mode structure agrees with the corresponding structure shown in Ref.~\cite{ishizawa2009thermal}}
    \label{fig:ishizawa-cut}
\end{figure}

\begin{figure}[ht!]
    \centering
    \includegraphics[scale=0.45]{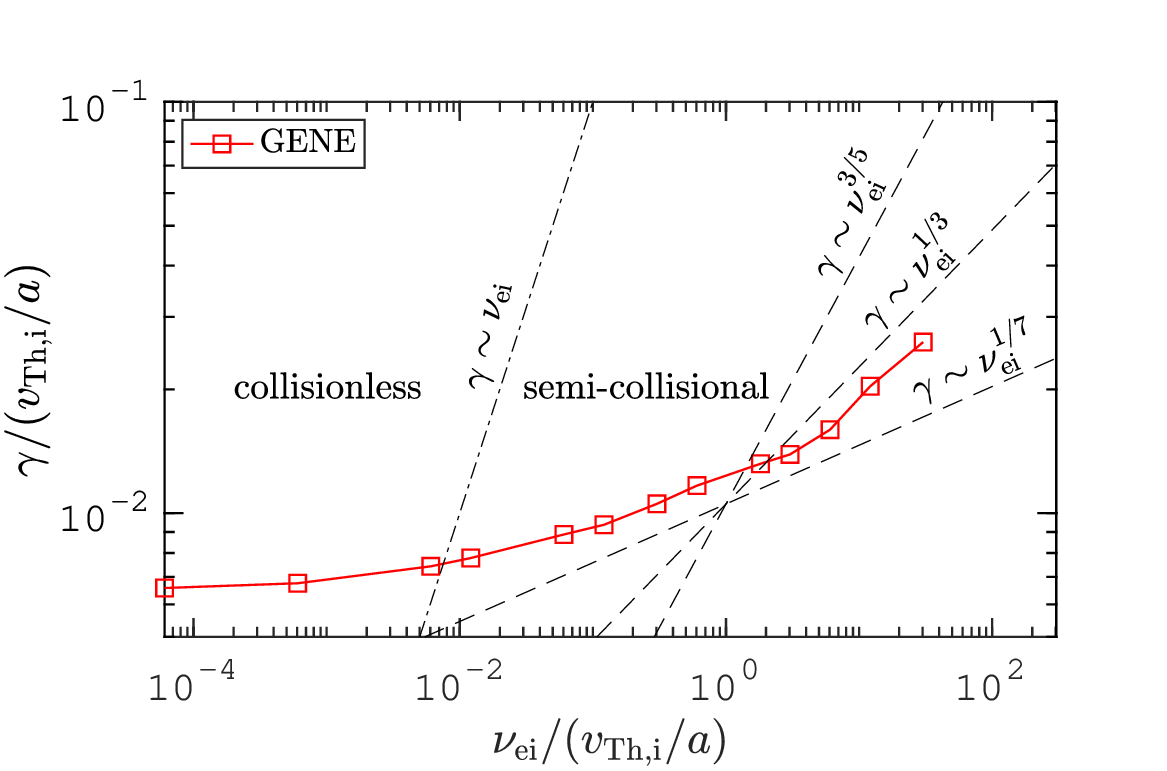}
    \caption{Growth rate of the $n=1$ tearing mode as a function of the electron-ion collision frequency for a benchmarking case with the two-fluid model, showing destabilization of the tearing modes with collisionality, obeying a scaling that follows theoretical predictions \cite{drake1977kinetic,porcelli1991collisionless}.} 
    \label{fig:ishizawaCollScan}
\end{figure}

\begin{figure}[ht!]
    \centering
    \includegraphics[scale=0.45]{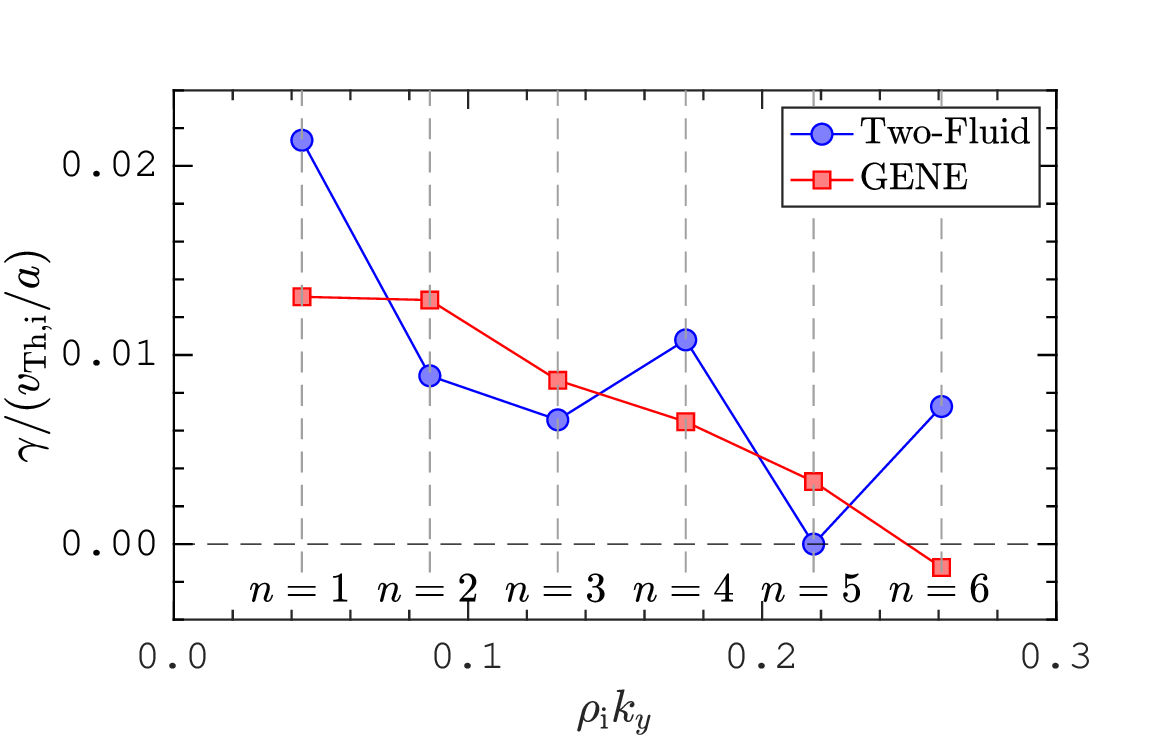}
    \caption{Growth rate spectrum of tearing mode obtained from \textsc{Gene} (blue circles) and the two-fluid model (red square).}
    \label{fig:ishizawa_spectrum}
\end{figure}

The tearing-mode's radial mode structure produced in \textsc{Gene} is shown in the bottom panel of Fig.~\ref{fig:ishizawaprof}, highlighting the characteristics of the $m/n$ = 2/1 mode and the subdominant structure of the $m/n=3/1$ mode. Fig.~\ref{fig:ishizawa-cut} shows the poloidal cross-section with a 2/1 island feature, closely matching the corresponding result shown in Ref.~\cite{ishizawa2009thermal}. 
Fig.~\ref{fig:ishizawaCollScan} shows a collisionality scan that demonstrates destabilization of tearing modes by collsionality. The growth rate of the $n=1$ mode scales as the theoretical prediction \cite{drake1977kinetic, porcelli1991collisionless}.
The growth rate spectrum is shown in Fig.~\ref{fig:ishizawa_spectrum}. Although the growth rate for each toroidal mode number $n$ does not exactly trace the points obtained from the fluid model, they agree within the uncertainty implied by the scatter of the data points of the fluid model. 

These benchmarking studies demonstrate that the modified \textsc{Gene} code is able to effectively model global tearing. This modeling capability enables the quantitative, bi-directional, and self-consistent study of multi-scale interactions between tearing modes and microinstabilities.
\section{Tearing Modes in the RFP}
Verification of the modified global \textsc{Gene} code against various computational models, as demonstrated in the preceding section, indicates the capability to accurately reproduce linear tearing modes in the RFP. Furthermore, it provides confidence that the code will serve as a valuable tool for investigating the nonlinear evolution of tearing modes, TEMs, and multi-scale interactions within the RFP system. Analysis of multi-scale interactions is a major undertaking and initial efforts will be presented in a separate paper. In this section, simulation of linear tearing modes for an RFP equilibrium will be shown and analyzed, followed by simulation of nonlinear tearing modes and their saturation.

RFPs are characterized by toroidal ($B_\varphi$) and poloidal ($B_\theta$) magnetic-field components that are of comparable magnitude. As a result, the safety factor in RFPs is much lower than unity. Specifically, at the magnetic axis, the safety factor value is typically given by $q(r_a=0)\sim a/(2R)$. As we move towards the plasma edge, the magnitude of the toroidal magnetic field gradually decreases, in standard RFP discharges eventually reversing its direction. Consequently, $q$ decreases with minor radius and becomes negative at the edge, which is in contrast to tokamaks where $q$ increases toward the edge. Both the toroidal and poloidal magnetic-field components in RFPs vary strongly with minor radius, which leads to significant magnetic shear. Combined with the low value of $q$, this enables the formation of low-order rational surfaces, making the plasma vulnerable to tearing modes \cite{marrelli2021reversed}.

\begin{figure}[ht!]
    \centering
    \includegraphics[scale=0.45]{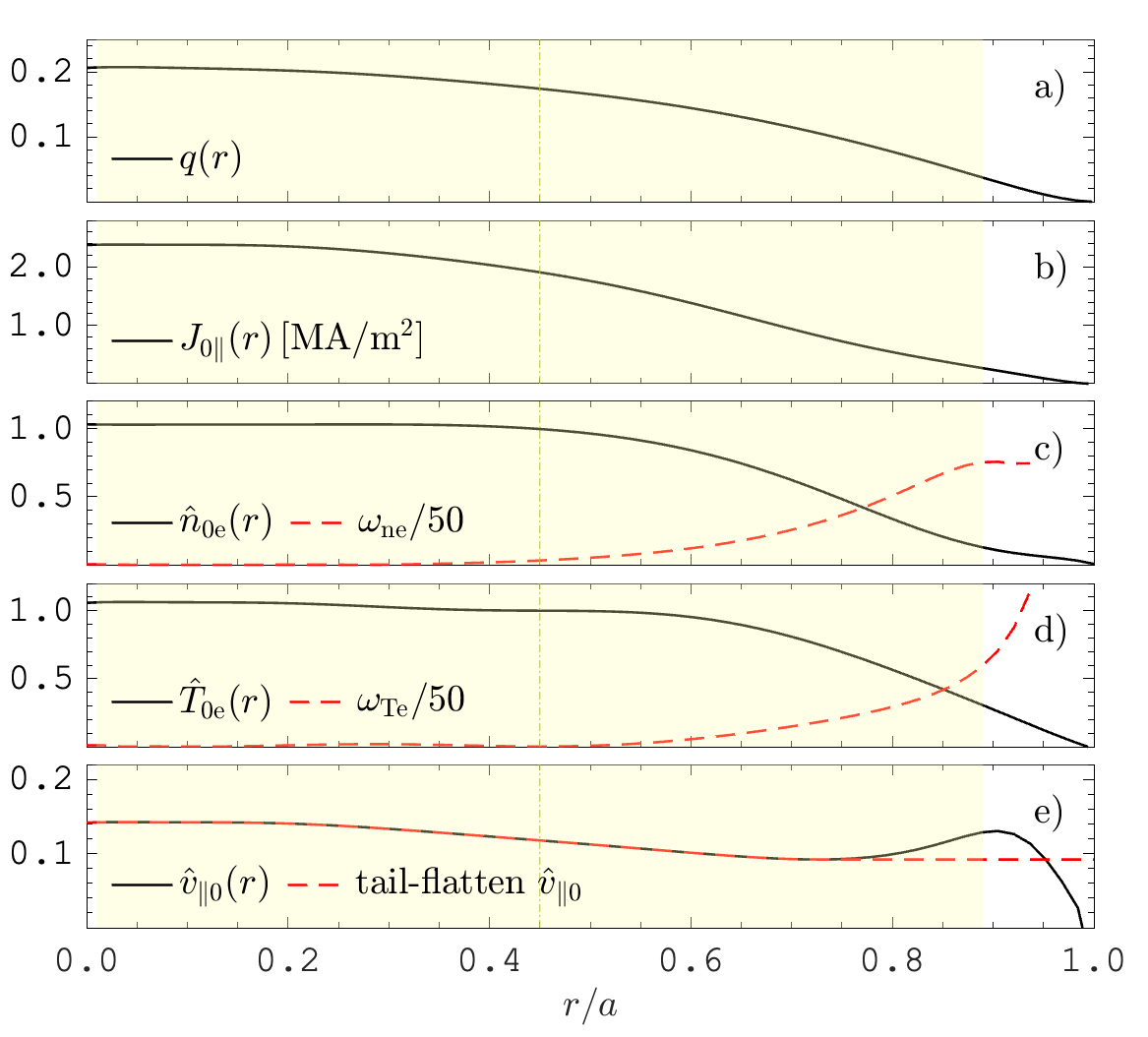}
    \caption{Equilibrium profiles of an $F=0$ discharge obtained from the MSTFit reconstruction code. The simulation domain is in the yellow area $r_a=[0.005,0.885]$ with the center at $r_a=0.45$.} 
    \label{rfp_profile}
\end{figure}

Previous numerical investigations of RFP plasmas were carried out using the nonlinear single-fluid visco-resistive MHD code DEBS \cite{schnack1987semi}. The code successfully modeled linear tearing modes and subsequent nonlinear evolution, agreeing with experimental observations, and enhancing our understanding of critical phenomena such as the dynamo effect, sawtooth oscillations, and nonlinear tearing saturation. Of particular significance is its capacity to include the reversal surface and $m=0$ modes in the analysis, which is a significant mediator of the energy cascade \cite{ho1991nonlinear, schnack1979nonlinear}. Another computational tool in the study of RFP tearing phenomena is NIMROD, a nonlinear extended MHD code \cite{sovinec2003nimrod}. NIMROD's unique feature lies in its ability to self-consistently account for drift effects and profile relaxation, enabling the investigation of fluctuation-induced transport impacting the plasma equilibrium \cite{king2012first, gupta2023pressure}. The code accounts for profile evolution and relaxation similar to experiments. In contrast, \textsc{Gene} offers distinct advantages by incorporating drifts, kinetic microturbulence effects, and zonal flows in the computation. This feature enables the study of multi-scale interactions. It is also important to note that \textsc{Gene} is unable to include the reversal surface in its simulation domain, as its coordinate system is not well-defined at this radius. Thus, the present study focuses on a non-reversed discharge, referred to as an $F=0$ discharge, where $F=B_\varphi(a)/\langle B_\theta\rangle$ is the reversal parameter. We obtained semi-analytical equilibrium profiles of the RFP from the MSTFit reconstruction code \cite{anderson2003equilibrium}, which solves the Grad-Shafranov equation using experimental data from the Madison Symmetric Torus (MST), discharge \#1130819032. In Fig.~\ref{rfp_profile}a, the $q$-profile is presented, with the reversal surface located at the plasma edge. This $F=0$ discharge avoids encountering the reversal surface with $q=0$. Unlike reversed discharges, where the turbulent cascade is mediated by $m=0$ modes, the specific nonlinear saturation mechanism for tearing modes in $F=0$ discharges has not yet been identified.

 The total current profile, shown in Fig.~\ref{rfp_profile}b, is utilized to calculate the shifted velocity (see Fig.~\ref{rfp_profile}e) for the shifted Maxwellian distribution function, as outlined in Eq.~(\ref{vshifted}). To mitigate numerical instability near the edge, we introduce a flattened region in the shifted velocity. This adjustment is expected to have a negligible impact on the linear growth rates and structures of tearing modes. The same adjustment is applied to nonlinear simulations. It is important to note that the edge region displays significant density and temperature gradients, seen in Fig.~\ref{rfp_profile}c-d, which can give rise to microinstabilities, particularly the $\nabla n$-driven TEM \cite{williams2017turbulence, carmody2015microturbulence}.

The nominal plasma parameters are computed at the center of the computational domain, i.e., at $r/a=0.45$. The reference  $\beta_\mathrm{e}$ is found to be 0.69\%. The normalized gyroradius $\rho_\mathrm{i}^*$ is calculated to be $1.46\times10^{-2}$. Since deuterium was used in the experimental setup of MST, the mass ratio employed in the simulation is $m_\mathrm{e}/m_\mathrm{i}=2.73\times10^{-4}$. The electron temperature $T_\mathrm{0e}$ serves as a reference temperature, while $T_\mathrm{0i}$ is set at 40\% of $T_\mathrm{0e}$ \cite{chapman2002high}. Considering collisions, the simulation incorporates the Landau collision operator with a collision frequency of $\nu_\mathrm{ei}={0.242\,c_\mathrm{s}/R_0}$. Various other collision operators were tested and yielded comparable linear growth rates. To maintain numerical stability and prevent fluctuations near the boundaries, a Dirichlet boundary condition with the Krook damping operator is applied to the buffer zones, which occupy 5\% of the radial domain on both ends. Heat and particle sources are applied in these zones to prevent the fluctuating quantities from deviating strongly from equilibrium, which can violate the $\delta f$ approximation. The $n=5$ resonant surface is close to the magnetic axis, so the simulation domain is positioned as close as feasible to the axis to mitigate mode suppression by the boundary. Accordingly, the simulation conducted is within the range $r/a = [0.005,0.885]$. The resolution employed in this case involves grid point counts for radial and parallel displacements, and for the parallel velocity and magnetic moment, with values of $N_x=512$, $N_z=16$, $N_{v_\parallel}=32$, and $N_\mu=16$, respectively.
 
In light of the distinct magnetic geometry of the RFP compared to that of the tokamak, the standard concentric circular magnetic geometry often used in \textsc{Gene} is ill-suited for accurately representing its true geometry. To overcome this constraint, Carmody \textit{et al.} \cite{carmody2015microturbulence} proposed an alternative circular geometry that accounts for the specific characteristics of RFPs. This modified approach, known as adjusted circular geometry, employs circular flux surface cross sections while incorporating a polynomial fit factor $f(r/R_0)$ to precisely capture the spatial variations of the total magnetic field across the radial extent. Specifically,
\begin{equation}
    \mathbf{B}=\frac{R_0B_0}{R}|q|f(r/R_0)\left(\mathbf{e}_\varphi+\mathbf{e}_\theta \frac{r}{R_0\bar{q}}\right),
\end{equation}
where $R_0$ and $r$ are the major and minor radii at the axis and of the flux surface, respectively, $R=R_0+r\cos\theta$, while $B_0$ is the total background magnetic field on the axis, and $\theta$ is the poloidal angle. The safety factor at the center of the simulation domain is denoted as $q$, and $\bar{q}=q\sqrt{1-(r/R_0)^2}$. The polynomial fit $f(r/R_0)$ allows consistency for the toroidal field when $r$ approaches the edge of the plasma. The model has been benchmarked and validated in studying microturbulence in the RFP \cite{carmody2015microturbulence}, and we modified and implemented this geometry for global simulations.

\subsection{Linear Analysis}
\begin{figure}[ht!]
    \centering
    \includegraphics[scale=0.42]{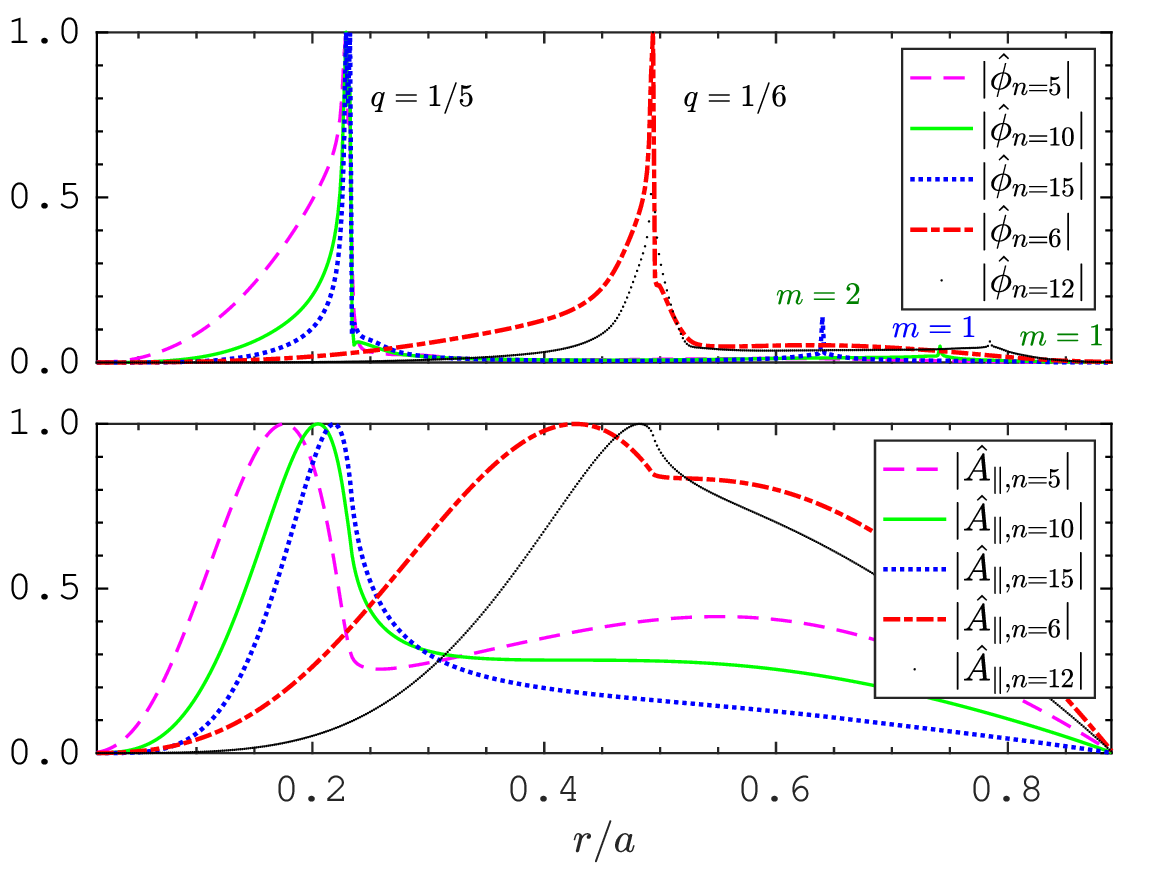}
    \caption{Radial mode structures of linear tearing modes. [Top] Electrostatic potential structures and [bottom] electromagnetic potential structures. The lowest-harmonic modes (e.g., $n=5$ and $n=6$) allow higher-harmonic modes (e.g., $n=10$ and $12$ of $n=5$, and $n=12$ of $n=6$) with different $m$ to resonate at the same rational surfaces. Nonlinearly, this allows different $n$ modes to couple.
    \label{rfp_structure}}
\end{figure}

\begin{figure}[ht!]
    \centering
    \includegraphics[scale=0.2625]{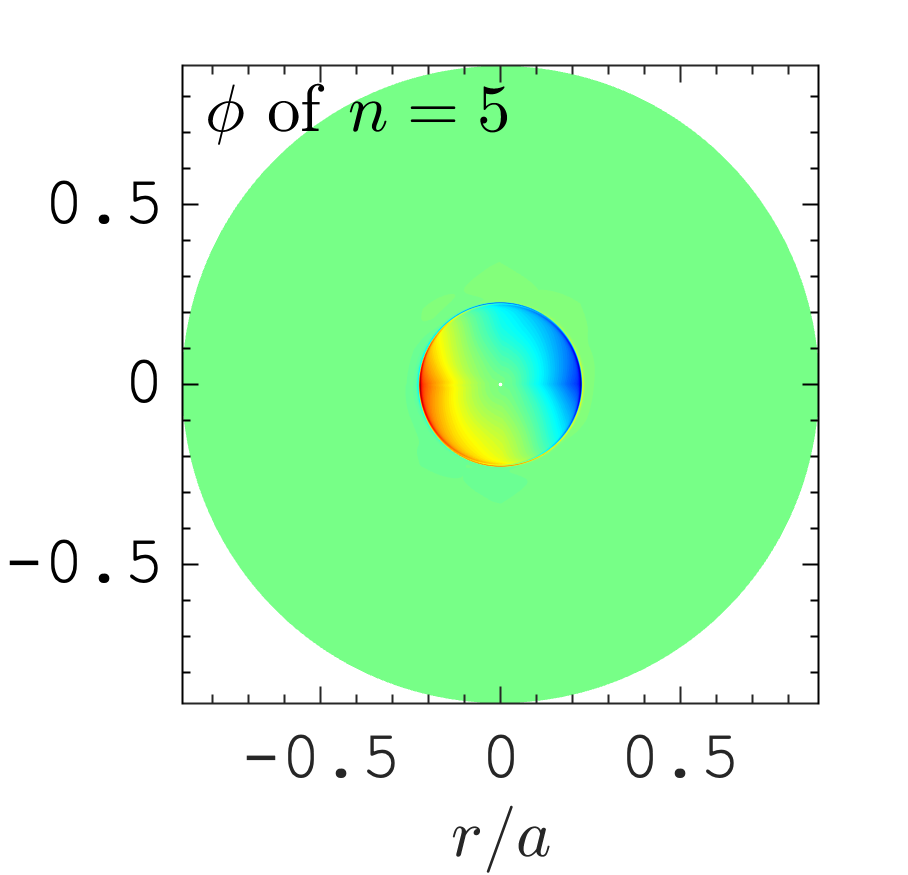}
    \includegraphics[scale=0.2625]{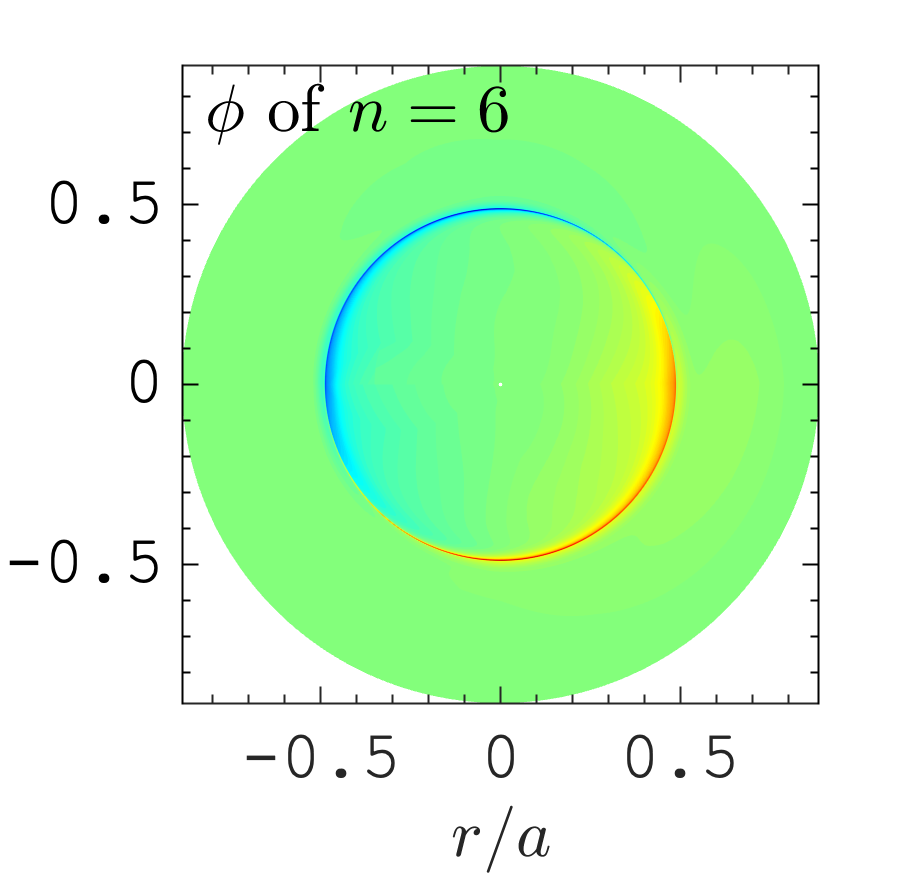}
    \caption{Poloidal cross-section of $n=5$ [left] and $n=6$ [right] electrostatic potential showing $m=1$ eigenmode structure on the $q=0.2$ and $q=0.16$ rational surfaces, respectively.}
    \label{rfp_torus}
\end{figure}

Since the $q$-profile on the axis has a maximum value slightly above 0.2, this allows the lowest possible resonating $(m,n)=(1,5)$ tearing mode to occur at $q=0.2$. Linear simulations performed with $n\geq5$ are shown in Fig.~\ref{rfp_structure}, illustrating the radial structures of the electrostatic potential $\phi$ and the electromagnetic parallel vector potential $A_\parallel$ of the $n=5$ and $n=6$ modes, along with some of their corresponding higher harmonics. Mode structures extend throughout the radial domain and exhibit well-known characteristics similar to those found in the benchmarking cases, i.e., an abrupt change in $\phi$ and $A_\parallel$ at the rational surfaces. The poloidal cross-section of the electrostatic potential of the $n=5$ and $n=6$ modes is illustrated in Fig.~\ref{rfp_torus}, clearly exhibiting the presence of an $m=1$ island, which is indicative of a tearing mode.

\begin{figure}[ht!]
    \centering
    \includegraphics[scale=0.44]{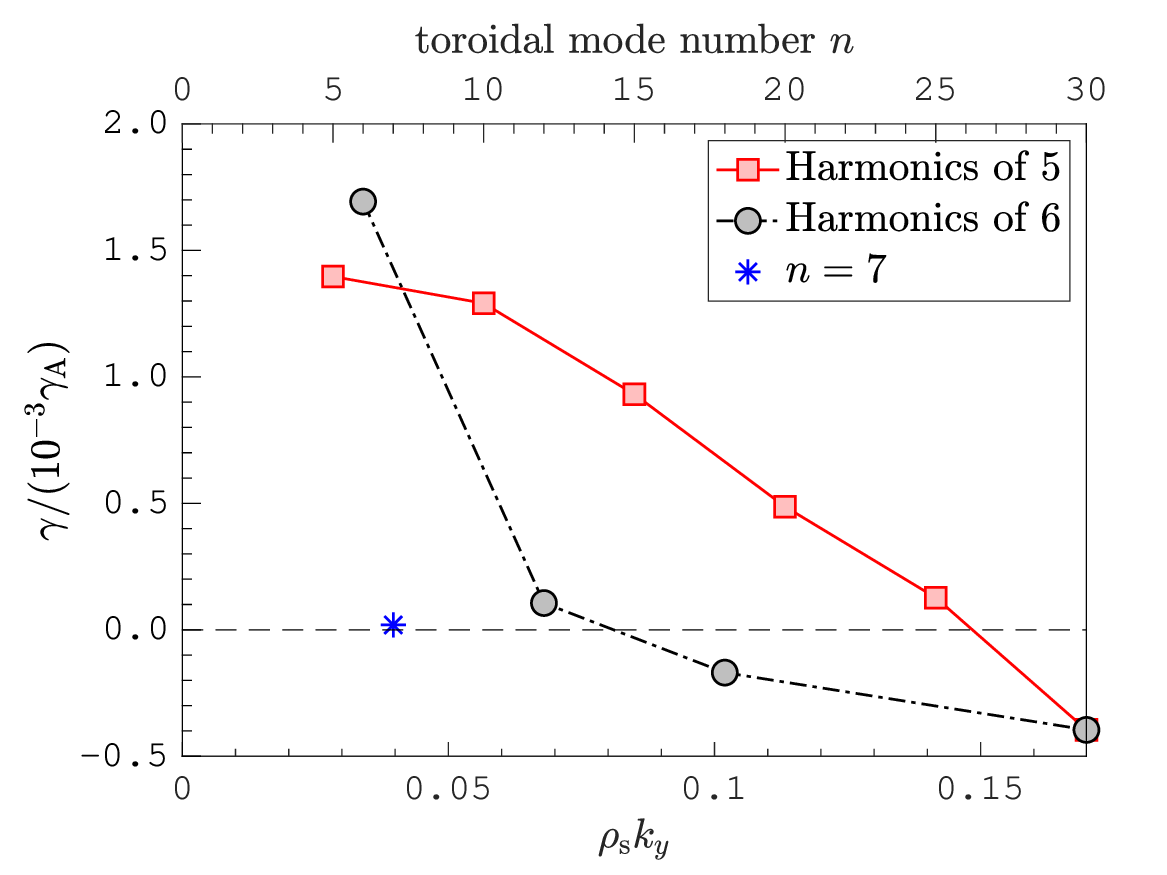}
    \caption{Linear tearing mode spectrum with respect to toroidal wavenumber. Note that $\gamma_\mathrm{A}=B_0/\sqrt{4\pi n_\mathrm{e} m_\mathrm{i}}/L_s$ and $L_s$ is the magnetic shear length defined as $L_s=r_s/\hat{s}=(r_s/R)(1/q^\prime(r_s))$ with $r_s$ the plasma radius at the rational surface of interest.} 
    \label{rfp_spectrum}
\end{figure}

Fig.~\ref{rfp_spectrum} shows the linear spectrum for different values of $n$. Among these modes, the ones with $n=5$ and $n=6$ exhibit linear instability, as they are directly driven by gradients in the background current. On the other hand, the mode with $n=7$ is marginally stable. Higher-order modes that are not resonant with $n=5$ and $n=6$, e.g., $n=8,\,9,\,...$, are found to be stable. Linear simulation in \textsc{Gene} allows for a single $n$ value but encompasses all attainable $m$ modes at different rational surfaces. For $n=5$, only the $m=1$ resonant surface lines withing the simulation domain, while for $n=10$, two $m$ modes can be resonance: $m=1$, which resonates near $r_a=0.74$, and $m=2$, which coincides with the resonance location of $(m,\,n)=(1,\,5)$, see Fig.~\ref{rfp_structure}. 

The spectrum in Fig.~\ref{rfp_spectrum} indicates that the growth rates of the dominant $n$ modes are approximately three orders of magnitude smaller than the Alfv\'en frequency $\gamma_\mathrm{A}$. This observation reflects the characteristic behavior of tearing modes, which exhibit significantly slower growth than an inverse Alfv\'en time. Additionally, our analysis reveals that the fundamental mode frequencies fall within the range of frequencies obtained from the MST experiment, specifically around 5-25 kHz \cite{ren2011experimental, sarff1993nonlinear}. This correspondence suggests that the tearing modes generated in our simulation are consistent with those observed in the experimental setting.

As $n$, and correspondingly $k_y$, increases, the growth rates gradually diminish, eventually stabilizing at approximately $k_y=0.15$. For $n\geq35$, the instabilities exhibit more electrostatic behavior, with $|\phi|/|A_\parallel|>10$, with the mode peak localized near the plasma edge. These particular instabilities are identified as $\nabla n$-TEM and will be the subject of comprehensive discussion in a separate publication.
~
\\

\begin{figure}[ht!]
    \centering
    \includegraphics[scale=0.45]{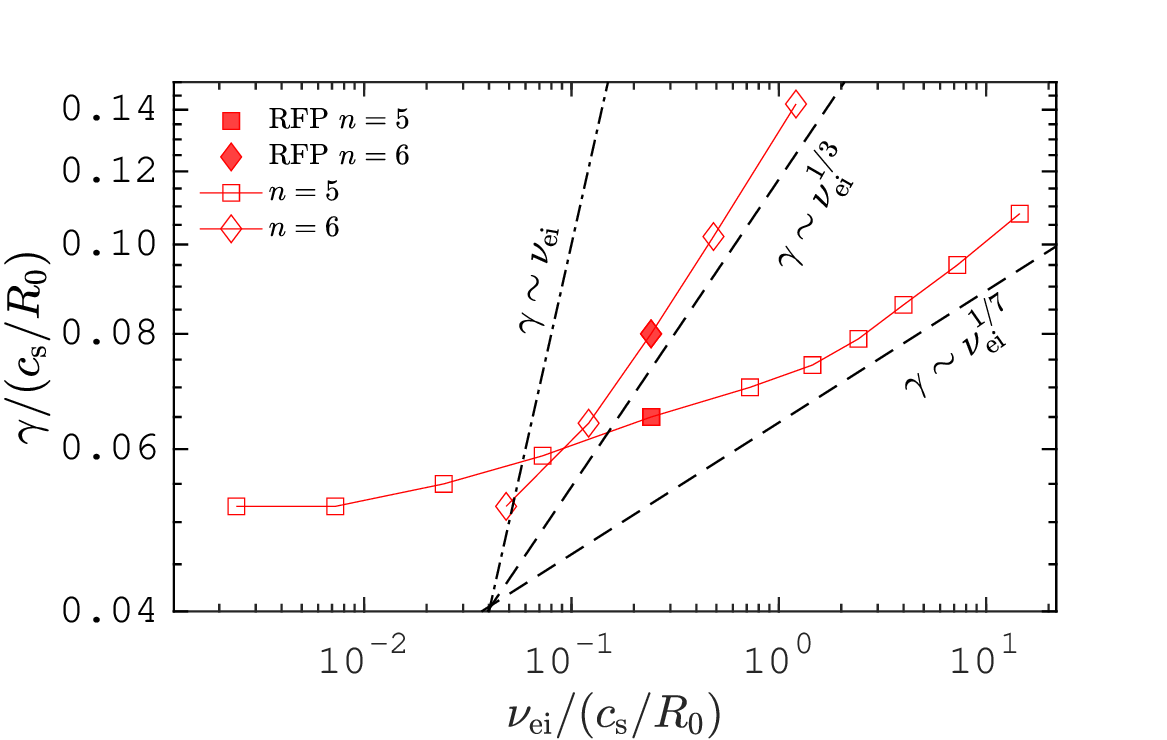}
    \caption{Collisionality scan for the $n=5$ tearing modes, showing two collisionality regimes: a collisionless regime where the growth rate asymptotically approaches a constant at very low collisionality, and a semi-collisional regime where the growth rates increase with a scaling of $\nu_\mathrm{ei}^{1/7}$. Semi-collisional regime of $n=6$ is also plotted, showing scaling of  $\nu_\mathrm{ei}^{1/3}$. A solid square and a solid diamond mark the growth rates of $n=5$ and $n=6$, respectively, at the collisionality representing MST-RFP, evaluated from given equilibria. } 
    \label{rfp_coll_scan}
\end{figure}

Consistent with the benchmark cases, the collisionality scans with $n=5$ and $n=6$ show destabilization at larger collision frequencies, indicating tearing instability for the RFP equilibrium. Fig.~\ref{rfp_coll_scan} shows the collisionality scan and indicates that there are collisionless and semi-collisional regimes at lower and higher collisionalities, respectively. The mode with $n=5$ shows a scaling $\gamma\propto\nu_\mathrm{ei}^{1/7}$, while the $n=6$ mode scales as $\gamma\propto\nu_\mathrm{ei}^{1/3}$ in the semi-collisional regime. Eq.~(\ref{esd}) shows that both $n=5$ and $n=6$ modes fall into a large-gyroradius regime in which the ion gyroradius is roughly on the same order as the resistive width characterized by the electron skin depth, $\rho_\mathrm{i}\sim\delta_\mathrm{e}$, and Ref.~\cite{porcelli1991collisionless} predicts that a tearing mode in such regime follows a $\gamma\propto\nu_\mathrm{ei}^{1/7}$ scaling. The behavior of $n=5$ mode, with $\rho_\mathrm{i}=2.4\delta_\mathrm{e}$, aligns with this prediction, unlike the $n=6$ mode, with $\rho_\mathrm{i}=2.6\delta_\mathrm{e}$, that exhibits a growth rate scaling $\gamma\propto\nu_\mathrm{ei}^{1/3}$ which matches predictions for the $\rho_\mathrm{i}<\delta_{\mathrm{e}}$ regime. The underlying reasons for this counter-intuitive behavior of the $n=6$ mode will be left for future work to resolve.

Linear simulations enable the identification and characterization of RFP tearing modes. These simulations offer valuable insights into the behavior of constituent modes and shed light on anticipated outcomes in nonlinear simulations. Notably, unstable tearing modes acquire energy from the background current gradient, thereby enabling their coupling with linearly stable modes.

\subsection{Nonlinear Tearing Evolution}
\begin{figure*}[ht]
    \centering
    \includegraphics[scale=0.48]{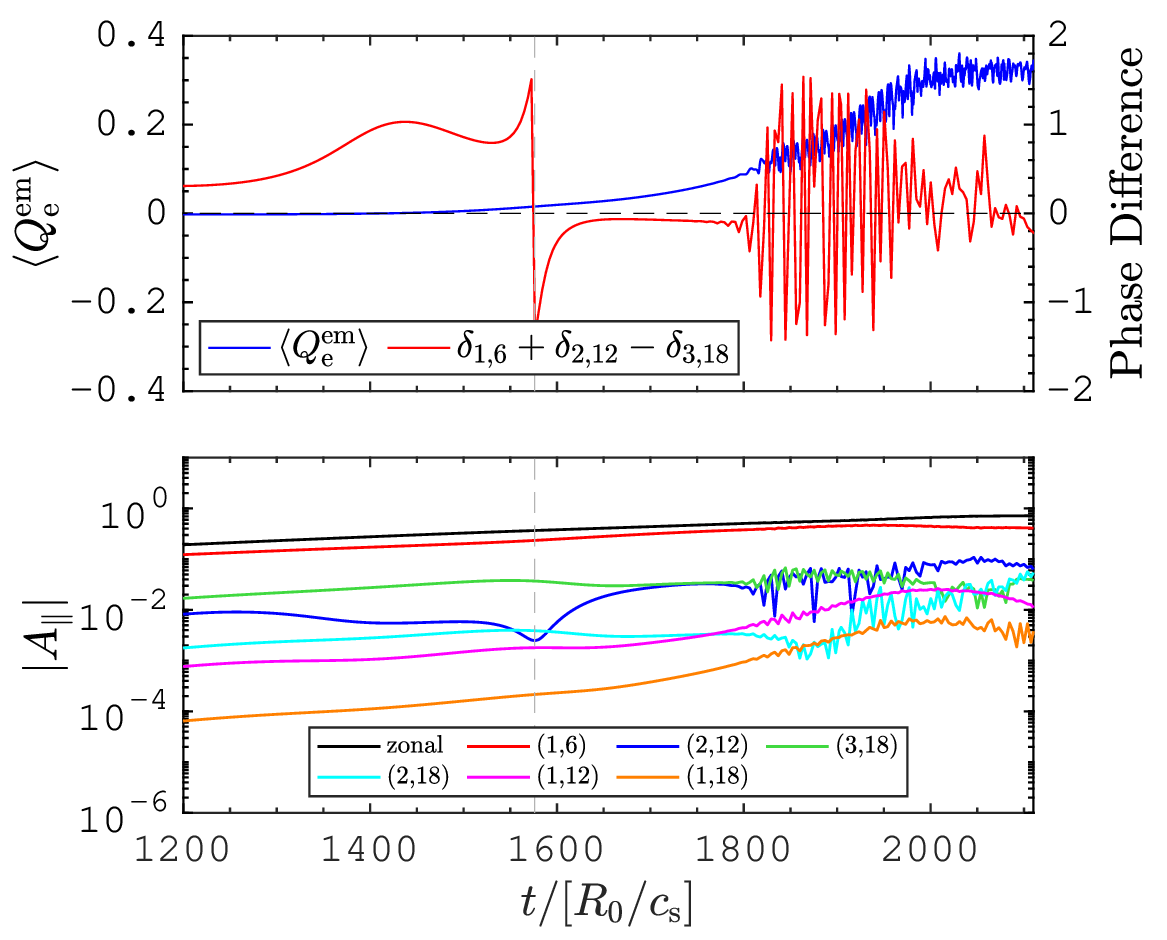}
    \caption{\label{fig:time_history} Time history of [top] the ensemble-averaged electromagnetic heat flux $\langle Q_\mathrm{e}^\mathrm{em}\rangle$ (blue line, left axis), phase difference $\Delta\delta = \delta_{1,6}+ \delta_{2,12}-\delta_{3,18}$ (red line, right axis), and [bottom] tearing-mode amplitudes available to resonate on rational surfaces in simulation domain. The plots show nonlinear dynamics with three different phases: a weak mode coupling phase ($t<1600$), a strong mode coupling ($1600<t<1850$), and a saturated phase with the quasi-steady heat flux ($t>1900$).}
\end{figure*}

To compare quantitatively with experiments, nonlinear simulations are required. The nonlinear simulation employs the same resolutions as those used in the converged linear calculations, with the exception of the number of the toroidal modes where $n=0,\,6,\,12,$ and $18$. The simulation is initiated at $t=0$ and eventually reaches a state where fluxes and mode amplitudes are saturated. The ensemble-averaged nonlinear electromagnetic electron heat flux $\langle Q_\mathrm{e}^\mathrm{em}\rangle$ plotted in the top panel of Fig.~\ref{fig:time_history} does not have a noticeable magnitude until the (exponentially) growing tearing mode becomes visible at approximately $t=1600$, as plotted in the bottom panel of Fig.~\ref{fig:time_history}.

In the initial phase of the simulation, prior to $t$$=$$1600$, the simulation reveals no significant mode coupling, as indicated by the small heat flux and wandering phase difference of the three modes resonant at the $q=1/6$ surface. As indicated by bispectral analysis \cite{sarff1993nonlinear, assadi1992measurement}, the excitation of one eigenmode through the interaction of other eigenmodes can result in a locked phase difference of zero; instead there is no locking during this early phase, indicating no clear coupling. To elaborate on this phenomenon, we consider the excitation of the electromagnetic vector potential of a tearing mode $A_{\parallel\,3,18}$ arising from the coupling of $A_{\parallel\,1,6}$ and $A_{\parallel\,2,12}$ as  $A_{\parallel,\,3,18}\propto A_{\parallel\,1,6} A_{\parallel\,2,12}$, where the first index in the subscript refers to the poloidal mode number $m$ and the second index refers to the toroidal mode number $n$. It is possible to write $A_{\parallel,\,m,n}$ as $|A_{\parallel,\,m,n}|\exp(i\delta_{m,n})$, with a spatial phase $\delta_{m,n}$, such that $|A_{\parallel,\,3,18}|\exp(i\delta_{3,8})\sim |A_{\parallel\,1,6} A_{\parallel\,2,12}|\exp(i\delta_{1,6}+i\delta_{2,12})$. Hence, if $A_{\parallel,\,3,18}$ is solely excited by the coupling of $A_{\parallel\,1,6}$ and $A_{\parallel\,2,12}$, the phase difference $\Delta\delta$$=$$\delta_{1,6}$$+$$\delta_{2,12}$$-$$\delta_{3,18}$ becomes zero. A fluctuation in phase difference can thus be interpreted as indicating weak coupling. The mode coupling strength is governed by fluctuation amplitude. Hence, when the flux is small, there is weak or no mode coupling and a wandering phase difference; when the flux is larger, three modes couple coherently; and when the flux is still larger, many modes are able to interact and scramble the phase difference of any three modes.

As the system progresses to $t>1600$, it becomes evident that the phases become locked, with the phase difference close to zero, showing strong mode coupling. Starting at approximately $t\approx1850$, multiple couplings occur simultaneously, causing an oscillation of mode amplitudes and a fluctuation phase difference around zero, while the flux increases. The couplings continue until the simulation reaches saturation around $t\approx2000$ by cascading. 

\begin{figure*}[ht]
    \centering
    \includegraphics[scale=0.60]{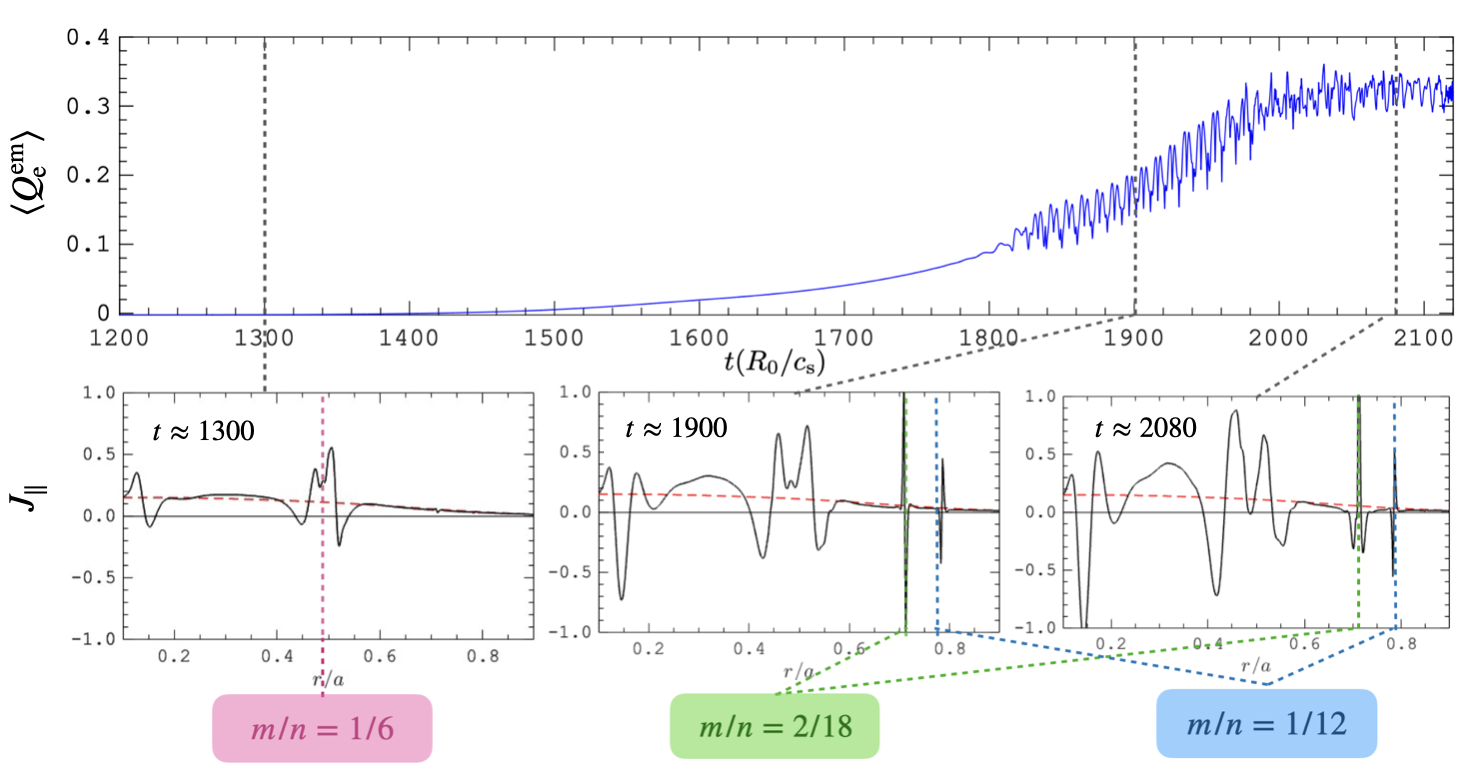}
    \caption{[Top] Time history of electromagnetic electron heat flux, and [bottom] radial profiles of fluctuating parallel velocity representing the parallel current density at different time snapshots. At $t\approx 1300$, only the strongest growing mode $m=1,\,n=6$ leads to a significant fluctuation level at the single rational surface. At $t\approx 1900$, modes $m=1$ and $m=2$ start to couple and lead to oscillations of the heat flux. At $t\approx 2080$, energy is transferred to $m=1,\,n=12$, as mediated by the $m=2$ mode. Fluctuations are less periodic at this stage as a result of mode couplings. The $m=1,\,n=12$ mode grows faster with a larger amplitude. Once the modes reach a quasi-stationary state, energy injected by the instability is balanced by dissipation at smaller scales.}
    \label{fig:nonlin}
\end{figure*}

Strong coupling leads to the onset of a cascade, allowing modes in the core to transfer energy to the edge \cite{sarff1993nonlinear}, i.e., a broadening of the tearing spectrum. As shown in Fig.~\ref{fig:nonlin}, the current density fluctuation at $t<1600$ illustrates a weak coupling with the presence of only the dominant tearing mode resonating at a rational surface with $q=1/6$. When coupling occurs at $t\approx1600$, the cascade begins, and multiple couplings follow at $t\approx 1850$, as observed in the middle panel of the second row of Fig.~\ref{fig:nonlin}. The fastest-growing mode starts to couple more with the linearly stable mode resonating at $q=1/9$, as observed in the amplitude of the current density on the surface. In this simulation, this corresponds to mode $m=2,\,n=18$, as the mode with $n=9$ is not present in the simulation. The coupling occurs between mode $m=2$ and mode $m=1$, with energy being transferred from mode $m=1$ to mode $m=2$. As the coupling involves a third mode with $(m,n)=(1,12)$, observed from the rising amplitude of current density at the last panel of the second row of Fig.~\ref{fig:nonlin}, fluctuations of the heat flux become more irregular. The interaction among all three modes leads to the saturation phase, where the fluctuation reaches a quasi-stationary state. By $t=2080$, the tearing modes are fully nonlinearly coupled, resulting in a higher amplitude of mode $(m,n)=(1,12)$. It is important to note that the $m=0$ mode does not play a significant role in energy mediation in this particular scenario, unlike the reversed discharges studied in Ref.~\cite{ho1991nonlinear}. In the present case, the $m=0$ mode is non-resonant and does not exhibit noticeable coupling with other modes in the system.

\section{Conclusions}
Tearing-mode fluctuations are significant players in the dynamics of RFP plasmas. It had been established from nonlinear simulations of $\nabla n$-TEM turbulence with the inclusion of an external magnetic field to model tearing modes, that tearing modes erode the microturbulent zonal flows and enhance the nonlinear electrostatic heat flux. Further studies of the interaction of slab ITG instability with a local code utilizing a shifted Maxwellian in the perturbed distribution had also indicated an effect of the tearing modes on the ITG turbulence level. However, these studies did not realistically model global tearing modes, with a lack of curvature effects, and did not account for the equilibrium current. The work described here is a modification of the global gyrokinetic code \textsc{Gene} to include current-gradient-drive terms arising from a shifted Maxwellian in the equilibrium distribution function. Implementation of the shifted Maxwellian distribution in the gyrokinetic equations enables realistic calculations of toroidal tearing mode instability and its nonlinear evolution self-consistently. The implementation includes modifying field equations and higher velocity moments to obtain physical observables under the shifted Maxwellian. 

The modified \textsc{Gene} code was benchmarked against different computational models. Comparisons were performed with the gyrokinetic PIC full-$f$ code ORB5, the $\delta f$ continuum code GKW, a reduced fluid model, and analytic theory. In comparison with the discrete gyrokinetic PIC code ORB5 utilizing gyrokinetic ions and electrons, scans over electron plasma $\beta$ and mass ratios show agreement between the two codes and consistency with theoretical scaling. A similar comparison was done using the equilibrium from Ref.~\cite{hornsby2015non}, where simulations of a linear tearing mode calculation using a continuous gyrokinetic code GKW. The growth rates obtained from \textsc{Gene} with different pressure gradients match well with results from GKW. Radial mode structures of the electrostatic and electromagnetic potentials and their peak ratios also match well. When the equilibria from the fluid simulation of Ref.~\cite{ishizawa2009thermal} were introduced for further testing, the modified \textsc{Gene} code produced mode structures that are in good agreement with the mode structures of the fluid calculation. The collisionality scaling and growth rate spectrum additionally confirm  the characteristics of global tearing modes. This establishes that the modified \textsc{Gene} code captures global tearing-mode dynamics in tokamak equilibria, and can be used to further study current-gradient-driven modes, especially with the presence of microturbulence.

The modified \textsc{Gene} code is also well equipped to study global tearing modes driven by the background current density in the RFP. While there exists a substantial body of research concerning benchmarking of microinstabilities in gyrokinetic codes, as well as MHD-scale instabilities in MHD codes, code-code comparisons using gyrokinetic simulations for MHD-scale modes have remained relatively limited. The present study represents results addressing this gap. The code is utilized to study the linear stability of a non-reversed $F=0$ discharge in MST. With the equilibrium modeled by the adjusted circular magnetic geometry of \textsc{Gene}, the simulation found that the most unstable linear tearing modes are $n=5$, which is close to the magnetic axis, and $n=6$, which is near the middle of the domain. Modes with $n=10$ and $n=12$ are also unstable. Higher harmonics eventually become stable, consistent with the stabilizing effect of larger $k_y$. Radial mode structures and collisionality scans confirm that the global instability observed in MST geometry is indeed a tearing instability. 

A nonlinear tearing mode simulation was also performed, treating the nonlinear interaction of multiples of $n=6$ modes. The nonlinear simulation reaches saturation by excitation and coupling of modes at different $n$ numbers, mediated by interaction with the $m=2$ mode, instead of the $m=0$ mediator commonly seen in simulations of reversed discharges. The mode coupling transfers energy to higher $n$, consistent with the well-known magnetic energy cascade in the RFP.

The results from nonlinear simulations provide an illustration of how large-scale tearing modes, active in the core, excite tearing modes at outer radii and thus potentially influence the small-scale TEMs near the edge. Namely, through mode coupling mediated by the $m=2$ component, low-order-$n$ modes couple and transfer energy to higher-$n$, stable tearing modes. The coupling and cascading continue through multiple stages, leading to the spreading of energy towards finer-scale tearing modes at higher $n$, and progressing towards the edge. As a result, this progression leads to the excitation of stable tearing modes near the plasma edge, demonstrating the capability of exerting an influence on the TEM region of MST.

This study of self-consistent linear and nonlinear tearing modes in the RFP lays the foundation for the investigation of multi-scale interactions with microturbulent fluctuations, which will be carried out in future work. Ongoing investigations of multi-scale interactions in the RFP, which will be reported separately, will inform both the magnetic turbulence properties of the RFP and interactions of magnetic perturbations with microturbulence in tokamak systems.

\section*{Acknowledgement}
The authors express sincere gratitude to J.K.~Anderson for providing the $F=0$ equilibrium profiles obtained from the MSTFit reconstruction code. Additionally, the authors extend their appreciation to C.R.~Sovinec for the valuable discussions regarding the numerical simulation of global tearing modes in the RFP. Special thanks are also due to Z.R.~Williams and the developers of the \textsc{Gene} code for offering assistance in implementing the global code for our study.

\end{document}